\documentclass[%
 reprint,
superscriptaddress,
 amsmath,amssymb,
 aps,
]{revtex4-1}
\usepackage{dcolumn}
\usepackage{diagbox}
\usepackage{tikz,colortbl}
\usepackage{graphicx}
\usepackage{dcolumn}
\usepackage{lipsum}
\usepackage{bm}
\usepackage{array}
\usepackage{siunitx}

\newcolumntype{C}[1]{>{\centering\arraybackslash}p{#1}}

\usepackage{tikz,colortbl}
\usetikzlibrary{calc}
\usepackage{zref-savepos}

\newcounter{NoTableEntry}
\renewcommand*{\theNoTableEntry}{NTE-\the\value{NoTableEntry}}

\newcommand*{\crossedout}{%
  \multicolumn{1}{@{}c@{}|}{%
    \stepcounter{NoTableEntry}%
    \vadjust pre{\zsavepos{\theNoTableEntry t}}
    \vadjust{\zsavepos{\theNoTableEntry b}}
    \zsavepos{\theNoTableEntry l}
    \hspace{0pt plus 1filll}%
    \zsavepos{\theNoTableEntry r}
    \tikz[overlay]{%
      \draw[black]
        let
          \n{llx}={\zposx{\theNoTableEntry l}sp-\zposx{\theNoTableEntry r}sp},
          \n{urx}={0},
          \n{lly}={\zposy{\theNoTableEntry b}sp-\zposy{\theNoTableEntry r}sp},
          \n{ury}={\zposy{\theNoTableEntry t}sp-\zposy{\theNoTableEntry r}sp}
        in
        (\n{llx}, \n{lly}) -- (\n{urx}, \n{ury})
        (\n{llx}, \n{ury}) -- (\n{urx}, \n{lly})
      ;
    }%
  }%
}

\begin{document}


\title{Fast general two- and three-body interatomic potential}

\author{Sergey Pozdnyakov}
\affiliation{Skolkovo Institute of Science and Technology Bolshoy Boulevard 30, bld. 1, 121205 Moscow, Russia}
\author{Artem R. Oganov}
\affiliation{Skolkovo Institute of Science and Technology Bolshoy Boulevard 30, bld. 1, 121205 Moscow, Russia}
\author{Efim Mazhnik}
\affiliation{Skolkovo Institute of Science and Technology Bolshoy Boulevard 30, bld. 1, 121205 Moscow, Russia}
\author{Arslan Mazitov}
\affiliation{Dukhov Research Institute of Automatics (VNIIA), Moscow 127055, Russian Federation}
\affiliation{Moscow Institute of Physics and Technology, 141700, 9 Institutsky lane, Dolgoprudny, Russian Federation}

\author{Ivan Kruglov}
\affiliation{Dukhov Research Institute of Automatics (VNIIA), Moscow 127055, Russian Federation}
\affiliation{Moscow Institute of Physics and Technology, 141700, 9 Institutsky lane, Dolgoprudny, Russian Federation}

\begin{abstract}
We introduce a new class of machine learning interatomic potentials---fast General Two- and Three-body Potential (GTTP), which is as fast as conventional empirical potentials and require computational time that remains constant with increasing fitting flexibility. GTTP does not contain any assumptions about the functional form of two- and three-body interactions. These interactions can be modeled arbitrarily accurately, potentially by thousands of parameters not affecting resulting computational cost. Time complexity is O(1) per every considered pair or triple of atoms. The fitting procedure is reduced to simple linear regression on \textit{ab initio} calculated energies and forces and leads to effective two- and three-body potential, reproducing quantum many-body interactions as accurately as possible. Our potential can be made continuously differentiable any number of times at the expense of increased computational time. We made a number of performance tests on \mbox{one-,} two- and three-component systems. The flexibility of the introduced approach makes the potential transferable in terms of size and type of atomic systems. We show that trained on randomly generated structures with just 8 atoms in the unit cell, it significantly outperforms common empirical interatomic potentials in the study of large systems, such as grain boundaries in polycrystalline materials.
\end{abstract}

\pacs{Valid PACS appear here}
\maketitle


\section{\label{sec:level1}Introduction}

In computational chemistry, the majority of calculations are performed within Born-Oppenheimer approximation \cite{Born_Op}, which states that the motion of atomic nuclei and electrons can be decoupled. Within this approximation, the potential energy of a system is completely defined by atomic positions, their types, and the total number of electrons in the system. Thus, the concept of potential energy surface (PES) is introduced as the functional dependence of the potential energy on the atomic positions. At each point, PES can be calculated by performing \textit{ab initio} electronic structure calculations, where atomic positions are considered as the parameters of the electronic Hamiltonian. But such calculations are computationally very demanding, and simpler methods are typically used. One such method is density functional theory (DFT) \cite{hohenberg1964, kohn1965}, which significantly reduces the parameter space by introducing the charge density. Another example is the tight binding (TB) model \cite{slater1954}, where the exact Hamiltonian is replaced by a parametrized matrix. Although these methods, especially DFT, remain quite accurate in many applications, they are still very computationally demanding, and thus it is hardly possible to use them for systems with more than several hundred atoms.

One possible way around this problem is to use conventional empirical interatomic potentials. In this approach, some fixed functional form with a few adjustable parameters is used for linking the potential energy and atomic positions. Such potentials are orders of magnitude faster, but their accuracy is limited, and for each type of compound, a different analytical form is needed. For example, different properties of metals are often modeled with the embedded atom method \cite{daw1984embedded}, modified embedded atom method \cite{baskes1992modified}, or angular-dependent potentials \cite{mishin2006angular}. Organic compounds are usually simulated with AMBER, CHARMM, or other force fields (a good review can be found in Ref. \cite{ponder2003force}). Different chemical processes and reactions, polymerization, and isomerization can be studied with a reactive force field (ReaxFF) \cite{van2001reaxff}.

Another way is becoming increasingly popular nowadays - machine learning potentials. Regression problem 
is one of the standard problems of machine learning. Examples vary from the prediction of age by photo \cite{age} to the prediction of the number of comments a blog post will receive based on its features \cite{feedback}. The approximation of the PES can also be formulated as a regression problem, and the general scheme is the following: first, energies and forces are calculated by \textit{ab initio} methods for some set of structures. Next, this dataset is used to fit some machine learning model, and after that, it can be used to efficiently and accurately predict energies and forces for new structures. A number of machine learning potentials were recently developed based on neural networks \cite{behler2007generalized, behler2011atom, behler2016perspective, BehlewWater, NNsodium1, NNsodium2, NNsilicon, ChargeNN, dolgirev2016machine}, gaussian regression \cite{bartok2010gaussian, bartok2013machine, bartok2013representing}, linear regression \cite{shapeev2016moment, podryabinkin2016active, kruglov2017energy, li2015molecular} and other approaches \cite{delta, MLIndia, yao2017many, mueller2016machine}. 

Thereby, conventional empirical potentials are the fastest, but their accuracy is limited. Electronic structure calculations have the best accuracy, but they are computationally very demanding. Machine learning potentials represent a compromise between these two approaches. 

In this paper, we report a general two- and three-body machine learning potential, which is as fast as conventional empirical potentials and, at the same time, is much more flexible.

The manuscript is structured as follows. In Section~\ref{sec:level2} we describe the methodology of the presented two- and three-body potential. Section~\ref{sec:comparison} contains a theoretical comparison with the other interatomic potentials. The new class of atomic invariant descriptors is introduced in  Section~\ref{sec:descriptors}. Section~\ref{sec:more_general} contains a generalization of the parametrization of the potential. In Section~\ref{sec:results} we report numerical experiments checking the effect of all hyperparameters, performance summary, computational cost, and extraction of chemically interpretable information from raw DFT calculations.

\section{\label{sec:level2}General two- and three-body potential}

The real quantum interactions between the atoms in a chemical system can not be reduced to two- and three-body terms. But in most cases, the main contribution to the energy variance can be ascribed to two- and three-body interactions. So we decided to focus on them and construct a model which is able to reproduce arbitrary two- and three-body interactions, at the same time being computationally efficient. 

For the sake of simplicity, subsequent paragraphs contain a description of the potential for the case of a single atomic type. The generalization for multiple atomic species is described later.

In two- and three-body interactions approximation, the energy of the system (except additive constant) is given by:
\begin{equation} 
\label{eq:energy0} E = \sum\limits_{i < j} E_2(\vec{r}_i, \vec{r}_j) + \sum\limits_{i < j < k} E_3(\vec{r}_i, \vec{r}_j, \vec{r}_k)\text{,}  \end{equation} where $i$, $j$ and $k$ runs over all atoms in the system, $\vec{r}$ are the positions of corresponding atoms, $E_2$ and $E_3$ are the energies of pair and triple interactions, respectively.

A pair of atoms has one degree of freedom - the distance between them, while a triple of atoms has three degrees of freedom, which we decided to choose as three sides of the corresponding triangle. Thus, Eq.~\ref{eq:energy0} can be rewritten as 

\begin{align}
E &= \sum\limits_{i < j} \varphi_2 (\mid\vec{r}_i -  \vec{r}_j\mid)  \nonumber \\&+  \sum\limits_{i < j < k} \varphi_3(\mid\vec{r}_i -  \vec{r}_j\mid, \mid\vec{r}_i -  \vec{r}_k\mid, \mid\vec{r}_j -  \vec{r}_k\mid) \text{,} \label{eq:energy1}
\end{align}
where $\varphi_2$ and $\varphi_3$ are one- and three-dimensional functions, which determine two- and three-body potentials. 
The summation in Eq.~\ref{eq:energy1} scales as $O(N^3)$, where $N$ is the number of atoms in the system, which is unacceptable. Thus, two cut-off radii $R_{cut}^2$ and $R_{cut}^3$ are introduced to discard long-range interactions. Now the summation in the first term is performed through only such pairs of atoms, where mutual distance is less than $R_{cut}^2$. Set of such pairs we will denote as $P(R_{\text{cut}}^2)$. Summation in the second term we implemented in two variants---in the first one summation is performed over triples of atoms, where every side of the corresponding triangle does not exceed $R_{\text{cut}}^3$, and in the second over triples of atoms, where at least two sides do not exceed $R_{\text{cut}}^3$. Sets of proper triples we will denote as $T(R_{\text{cut}}^3)$ for both variants. After such cutting, the complexity of the potential becomes the desired $O(N)$. The values of $R_{\text{cut}}^2$ and $R_{\text{cut}}^3$ represent the tradeoff between speed and accuracy. The higher $R_{\text{cut}}^2$ and $R_{\text{cut}}^3$, the more accurate and slower the potential is. For different chemical systems, the best compromise between time and accuracy can be achieved with different variants of triples cutting. Thus, these two implemented ways to do it provide additional flexibility. 

So, to determine the two- and three-body potential, one needs to determine functions $\varphi_2$ and $\varphi_3$ on finite domains. We decided to parametrize them in the form of piece-wise polynomials on an equidistant grid. But the arbitrary coefficients for these polynomials are not suitable because the resulting PES approximation should obey certain continuity properties. For example, interatomic potentials are often used in molecular dynamics, where forces---derivatives of the energy with respect to atomic positions, are needed. Thus, PES approximation, and therefore functions $\varphi_2$ and $\varphi_3$ should be continuously differentiable. This means that one needs to impose additional stitching conditions on polynomial coefficients.

While the most prevalent demand for the potential is to be once continuously differentiable, sometimes a need for greater smoothness can arise. Our framework supports constructing arbitrarily many times continuously differentiable potentials. 

Domain for the $\varphi_2$ is the interval from some $S_2 \geq 0$ to $R_{\text{cut}}^2$. It makes sense to choose $S_2 \neq 0$ because in all chemical systems there exists some minimal distance such that the probability of two atoms being closer is vanishingly small. In practice, after fitting the potential, we continue $\varphi_2$ from $S_2$ or even from some $C_2 > S_2$ to zero in accordance with the required smoothness in such a way that it tends to infinity at zero. This is needed to correctly handle such very rare situations as the ones in molecular dynamics when two atoms might come extremely close to each other. We use the equidistant grid containing $Q_2 + 1$ vertices, $Q_2 - 1$ inner vertices, and thus $Q_2$ \text{intervals}, which are enumerated from $0$.

If constructed potential is required to be $k - 1$ times continuously differentiable, we use polynomials of order $k$ and $\varphi_2(r)$ is given by:

\begin{equation} \label{eq:old_parametrization}
\varphi_2(r) = \sum_{l= 0}^{k} a_p^l  r^l \text{,}
\end{equation}
where $a_p^l$ is $l$-th coefficient of the polynomial on the $p$-th interval and $p = \lfloor Q_2 \frac{r - S_2}{R_{\text{cut}}^2 - S_2} \rfloor$ is the index of the interval to which $r$ belongs.

The values of polynomials and their $k - 1$ derivatives should match in all inner vertices. In addition, the value of the last polynomial and its $k - 1$  derivatives at $R_{\text{cut}}^2$ should be equal to zero. Thus, arbitrary coefficients $a_p^l$ are not suitable.

\begin{figure}[h]
    \centering
    \includegraphics[width=8cm]{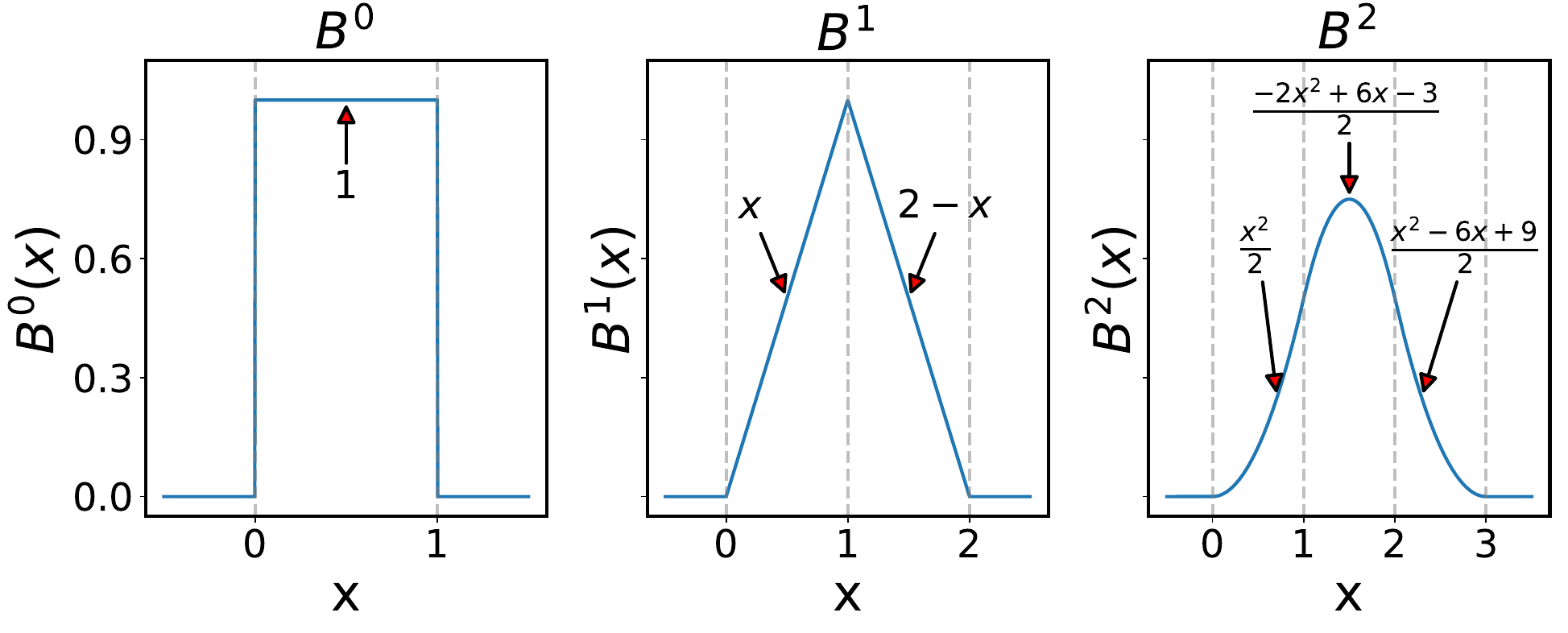}
    \caption{Cardinal B-splines of $0$, $1$ and $2$-nd order}
    \label{fig:Cardinal_B_splines}
\end{figure}

The way to ensure these stitching conditions is to use parametrization with cardinal B-splines, which are a special case of B-splines when the grid is equidistant. Cardinal B-spline of $k$-th order is the $k - 1$ times continuously differentiable (when $k > 1$) piece-wise polynomial function of $k$-th order on each interval, whose support consists of $k + 1$ equidistant \text{intervals}. Cardinal B-splines of $0$, $1$ and $2$-nd order are shown  in Fig.  \ref{fig:Cardinal_B_splines}

Cardinal B-splines of arbitrary order can be calculated using the Cox-de Boor recursion formula \cite{deboorcox1, deboorcox2}.

\begin{figure}[h]
    \centering
    \includegraphics[width=8cm]{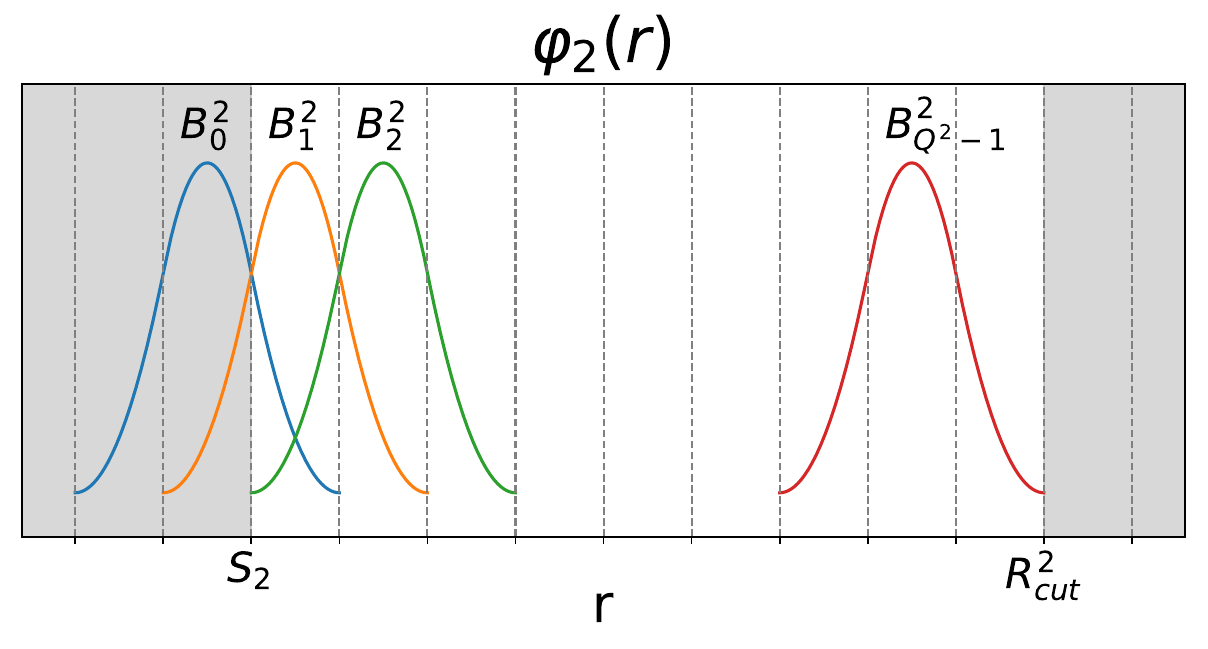}
    \caption{Cardinal B-splines parametrization}
    \label{fig:parametrization}
\end{figure}

The new parametrization for $\varphi_2(r)$ is:
\begin{equation} \label{eq:new_parametrization}
    \varphi_2(r) = \sum_{m = 0}^{Q_2 - 1} c_m B_m^k(r) \text{,}
\end{equation}
where $c_m$ are parametrization coefficients, $B_m^k(r)$ are cardinal B-splines of order $k$ and whose supports spread from $m - k$ to $m$-th  interval,  see Fig. \ref{fig:parametrization}

It is clear that any function in the form of Eq.~\ref{eq:new_parametrization} with arbitrary coefficients $c_m$ is a piece-wise polynomial and obeys necessary stitching conditions. Also, it can be shown \cite{deboor} that any function in the form of Eq.~\ref{eq:old_parametrization}, which obeys required stitching conditions, can be parametrized in the form of Eq.~\ref{eq:new_parametrization}.

Hyperparameter $k$ controls how many times $\varphi_2$ is continuously differentiable. But the greater this value, the higher the order of each polynomial and the higher are computational costs.

Now we will consider the three-dimensional $\varphi_3$ function, which determines three-body interactions. Its arguments are lengths of the sides of the triangle, which we denote as $r_1$, $r_2$, and $r_3$. The domain of $\varphi_3$ in the case of the first variant of triples cutting is the part of the cube $S_3 \leq r_1, r_2, r_3 \leq R_{\text{cut}}^3$, where $r_1$, $r_2$ and $r_3$ satisfy the triangle inequality. 

Similarly to $\varphi_2$, we introduce an equidistant grid and put $\varphi_3$ to be polynomial on each elementary cube.  
Thus, $\varphi_3$ is given by:

\begin{equation}
    \varphi_3(r_1, r_2, r_3) = \sum_{l_1, l_2, l_3 = 0} ^ {k}  b_{p_1, p_2, p_3} ^{l_1, l_2, l_3}  r_1 ^ {l_1}  r_2 ^ {l_2}  r_3 ^ {l_3} \text{,}
\end{equation} where $b_{p_1, p_2, p_3}^{l_1, l_2, l_3}$ are coefficients of the three-dimensional polynomial placed in the elementary cube with indices $p_1$, $p_2$, $p_3$, 
$p_\alpha = \lfloor Q_3 \frac{r_\alpha - S_3}{R_{\text{cut}}^3 - S_3} \rfloor$, $\alpha = 1, 2, 3$. 

As was stated earlier, arbitrary coefficients $b_{p_1, p_2, p_3}^{l_1, l_2, l_3}$ are not suitable, and thus three-dimensional cardinal B-splines parametrization is used. The three-dimensional cardinal B-spline is given by:
\begin{equation}
    B_{m_1, m_2, m_3}^k(r_1, r_2, r_3) = B_{m_1}^k(r_1) B_{m_2}^{k}(r_2) B_{m_3}^{k}(r_3) \text{.}
\end{equation}

The $\varphi_3$ function should be symmetric with respect to permutations of the sides of the triangle. Thus symmetric combinations of three-dimensional cardinal B-splines BS are used for the basis:

\begin{equation} \label{eq:symmetrization}
    BS_{m_1, m_2, m_3}^k(r_1, r_2, r_3) = \sum_{\alpha_1, \alpha_2, \alpha_3} B_{\alpha_1, \alpha_2, \alpha_3}^k (r_1, r_2, r_3) \text{,}
\end{equation}
where the summation is taken through all permutations of $m_1, m_2, m_3$.

So possible parametrization for $\varphi_3$ can be given as:
\begin{equation} \label{eq:initial_parametrization}
\begin{split} 
    &\varphi_3(r_1, r_2, r_3) = \\ &\sum_{ 0 \leq m_1 \leq m_2 \leq m_3 \leq Q_3 - 1} d_{m_1, m_2, m_3}  BS_{m_1, m_2, m_3}^k(r_1, r_2, r_3)  \text{.}
\end{split}
\end{equation}
This parametrization can be reduced because, due to triangle inequality, some terms in Eq.  \ref{eq:initial_parametrization} will never affect the energy. Thus, the final parametrization is:
\begin{equation}  \label{eq:final_parametrization}
\begin{split}
    &\varphi_3(r_1, r_2, r_3) = \\& \sum_{ \{m_1, m_2, m_3\} \in Z} d_{m_1, m_2, m_3}  BS_{m_1, m_2, m_3}^k(r_1, r_2, r_3) \text{,}
\end{split}
\end{equation}
where $Z$ is defined as subset of ${0 \leq m_1 \leq m_2 \leq m_3 \leq Q_3 - 1}$, which contains only such $\{m_1, m_2, m_3\}$ that there exist such $\{r_1, r_2, r_3\}$ satisfying triangles inequality that $BS_{m_1, m_2, m_3}^k(r_1, r_2, r_3) \neq 0$.

In the case of the second variant of triples cutting, the domain for $\varphi_3$ is more complex, but still, the parametrization can be done in a similar manner.

So the fitting process of two- and three-body potential is reduced to determining the coefficients $c_m$ and $d_{m_1, m_2, m_3}$. For this purpose, the functional dependence of the energy on these coefficients was investigated and turned out to be linear:
\begin{equation} \label{eq:energy_linear}
    E = \sum_{i = 0}^{Q_2 - 1} c_m D_m^2 + \sum_{ \{m_1, m_2, m_3\} \in Z} d_{m_1, m_2, m_3} D_{m_1, m_2, m_3}^3 \text{,}
\end{equation}
where $D_m^2 = \sum\limits_{<i, j> \in P(R_{\text{cut}}^2)} B_m^k(\mid\vec{r}_i -  \vec{r}_j\mid)$ and 
$D_{m_1, m_2, m_3}^3 = $
$$= \sum\limits_{<i, j, k> \in T(R_{\text{cut}}^3)} BS_{m_1, m_2, m_3}^k(\mid\vec{r}_i -  \vec{r}_j\mid, \mid\vec{r}_i -  \vec{r}_k\mid, \mid\vec{r}_j -  \vec{r}_k\mid)$$

Consequently, forces also depend linearly on the coefficients $c_m$ and $d_{m_1, m_2, m_3}$:
\begin{equation}
\begin{split}  \label{eq:forces_linear}
    F_{q_\alpha} &= -\frac{\partial E}{\partial r_{q_\alpha}}   =\sum_{m= 0}^{Q_2 - 1} c_m  (-\frac{\partial D_m^2} {\partial r_{q_\alpha}}) \\&+ 
    \sum_{  \{m_1, m_2, m_3\} \in Z} d_{m_1, m_2, m_3}  (-\frac{\partial D_{m_1, m_2, m_3}^3} {\partial r_{q_\alpha}})\cr\cr
\end{split}
\end{equation}

Thus, the fitting process is reduced to solving a linear regression problem, and the general scheme is the following:

For a given dataset, which contains structures and corresponding \textit{ab initio} calculated energies and forces, we

1) calculate values $D_m^2$, $D_{m_1, m_2, m_3}^3$, $ \frac{\partial D_m^2} {\partial r_{q_\alpha}}$ and $\frac{\partial D_{m_1, m_2, m_3}^3} {\partial r_{q_\alpha}}$ for every structure,

2) solve a joint linear regression problem, where input variables are values calculated at step 1, and target variables are energies and forces. The found coefficients of the linear model are $c_m$ and $d_{m_1, m_2, m_3}$,

3) convert $c_m$ and $d_{m_1, m_2, m_3}$ to coefficients $a_p^l$ and $b_{p_1, p_2, p_3} ^{l_1, l_2, l_3} $.

After this the potential is ready since coefficients $a_p^l$ and $b_{p_1, p_2, p_3} ^{l_1, l_2, l_3}$ completely determine two- and three-body potential.
In our implementation derivatives  $\frac{\partial D_m^2}{\partial r_{q_\alpha}}$ and $\frac{\partial D_{m_1, m_2, m_3}^3}{\partial r_{q_\alpha}}$ are calculated analytically. 

In the case of a multicomponent system, the energy is given by:
\begin{widetext}
\begin{equation} \label{eq:energy_multiple}
\begin{split}
&E = \sum\limits_{I \leq J} \sum\limits_{<i, j> \in P_{IJ}(R_{\text{cut}}^2)} \varphi_2^{I, J} (\mid\vec{r}_i -  \vec{r}_j\mid)  + \sum\limits_{I \leq J \leq K} \sum\limits_{<i, j, k> \in T_{I, J, K}(R_{\text{cut}}^3)} \varphi_3^{I, J, K}(\mid\vec{r}_i -  \vec{r}_j\mid, \mid\vec{r}_i -  \vec{r}_k\mid, \mid\vec{r}_j -  \vec{r}_k\mid) \text{,}
\end{split}
\end{equation}
\end{widetext}
where $I$, $J$, and $K$ run through atomic species, $P_{IJ}(R_{\text{cut}}^2)$ are the sets of atomic pairs, where atoms have types $I$ and $J$, $T_{I, J, K}(R_{\text{cut}}^3)$ are, analogously, sets of atomic triples, $\varphi_2^{I, J}$ and $\varphi_3^{I, J, K}$ are functions, which describe contributions to the energy from the pairs and triples with certain compositions.

If the total number of atomic species in the system is $N_t$, then the number of $\varphi_2^{I, J}$ and $\varphi_3^{I, J, K}$ functions is $\frac{N_t (N_t + 1)}{2}$ and $\frac{N_t (N_t + 1) (N_t + 2)}{6}$, respectively. The parametrization for all these functions is the same as discussed earlier for the case of a one-component system with the only difference that the symmetry for the $\varphi_3^{I, J, K}$ is applied only through triangles sides, which are equivalent with taken into account atomic species. In other words, if all $I$, $J$, and $K$ are the same, then the symmetry is applied through all 3 triangles' sides, and summation in Eq.  \ref{eq:symmetrization} contains 6 terms, if two of $I$, $J$ and $K$ are the same and the third is different, then the symmetry is applied only through 2 triangles sides and summation in Eq.  \ref{eq:symmetrization} contains 2 terms, and if all $I$, $J$ and $K$ are different, then symmetry is not applied, and summation in Eq.  \ref{eq:symmetrization} contains 1 term, or, equivalently, initial 3 dimensional cardinal B-splines are used as basis functions. We denote the corresponding symmetric combinations as $BS^k_{{IJK}_{m_1, m_2, m_3}}$. 

Also, for different symmetries, summation in Eq.  \ref{eq:final_parametrization} should be performed through different triples of indices, which we will denote as $Z_{IJK}$. Inequalities $m_\alpha \leq m_\beta$ should be satisfied only if $r_\alpha$ and $r_\beta$ are equivalent in a triangle constructed from atoms with types $I$, $J$ and $K$; as earlier, triangles inequality cutting should be performed.

Eventually, the Eq.~\ref{eq:energy_linear} and \ref{eq:forces_linear} transform into:
\begin{equation}
\begin{split}
    &E = \sum\limits_{I, J} \sum_{m= 0}^{Q_2 - 1} c_{IJ_m}  D_{IJ_m}^2   \\
    &+ \sum\limits_{I, J, K} \sum_{ \{m_1, m_2, m_3\} \in Z_{IJK}} d_{IJK_{m_1, m_2, m_3}}  D_{IJK_{m_1, m_2, m_3}}^3
\end{split}
\end{equation}
and 

\begin{equation}
\begin{split}
& F_{q_\alpha}  =  \sum\limits_{I, J} \sum_{m = 0}^{Q_2 - 1} c_{IJ_m}  (-\frac{\partial D_{IJ_m}^2} {\partial r_{q_\alpha}})  \\
    &+\sum\limits_{I, J, K} \sum_{  \{m_1, m_2, m_3\} \in Z_{IJK}} d_{IJK_{m_1, m_2, m_3}}  (-\frac{\partial D_{IJK_{m_1, m_2, m_3}}^3} {\partial r_{q_\alpha}}) \text{,}
\end{split}
\end{equation}
where $D_{IJ_m}^2 = \sum\limits_{<i, j> \in P_{IJ}(R^2_{\text{cut}})} B_m^k(\mid\vec{r}_i -  \vec{r}_j\mid) $  \newline
and 
$D_{IJK_{m_1,m_2,m_3}}^3 = $
$$=\kern-1em\sum\limits_{<i, j, k> \in T_{IJK}(R_{\text{cut}}^3)} \kern-2em BS_{IJK_{m_1,m_2, m_3}}^k(\mid\vec{r}_i -  \vec{r}_j\mid, \mid\vec{r}_i -  \vec{r}_k\mid, \mid\vec{r}_j -  \vec{r}_k\mid)$$
\newline
\newline
So single linear regression should be solved to simultaneously obtain all $c_{IJ}$ and $d_{IJK}$  coefficients and thus fit multicomponent two- and three-body potential. 

It is a well-known fact that any continuous one-dimensional function can be approximated on the segment arbitrarily close in the form of Eq.  \ref{eq:new_parametrization} by reducing grid spacing or, which is the same, increasing $Q_2$ \cite{deboor2}. The same also applies to the three-body potential. 

At the same time, complexity during the calculation of energies and forces does not depend on $Q_2$ and $Q_3$. Indeed, for every considered atomic pair or triple, the value of only one one- or three-dimensional polynomial of order $k$ or its derivative should be calculated. Computational costs per single atomic pair or triple increase with hyperparameter $k$, but it only relates to desired smoothness of the potential and does not control the fitting flexibility. In practice, we use $k = 2$ for all potentials in this work. In other words, the number of adjustable parameters does not affect computational time. It is especially beneficial in the case of multicomponent systems with a large number of atomic species where the number of these parameters can literally be thousands due to a large number of functions $\varphi_3^{I, J, K}$ and a large proportion of asymmetric or only partially symmetrical among them. In practice, the numbers of intervals of two- and three-body grids $Q_2$ and $Q_3$ are chosen long away in saturation area if the training dataset is big enough.

\section{\label{sec:comparison}comparison with other interatomic potentials}

The majority of existing conventional empirical potentials have a fixed functional form. Examples are Tersoff \cite{tersoff1988new},  Stillinger--Weber \cite{stillinger1986erratum}, and classical Lennard--Jones \cite{lennard} potentials. These potentials have a fixed number of adjustable parameters, so their accuracy is limited. Sometimes, the resulting functional form is constructed from a set of one-dimensional functions parametrized by splines. Examples are Lenosky \cite{lenosky2000highly}, and Zhang \cite{zhang2016modified}, where the three-body term in the modified embedded atom model (MEAM) is factorized as:
\begin{equation}
\psi(r_1, r_2, \theta) = f_1(r_1) f_2(r_2) f_3(\theta) \text{,}
\label{eq:factorization}
\end{equation}
where $f_1$, $f_2$ and $f_3$ are one-dimensional functions. This approach dramatically enriches the scope of functional forms it can parametrize, but it is clear that any three-dimensional function cannot be approximated arbitrarily close in the form of Eq.~\ref{eq:factorization}. 

On the other hand, machine learning potentials are much more flexible, but their computational time increases with fitting flexibility. For neural networks, for instance, both expressivity and the number of multiplications in forward pass depend on the number of neurons, and thus, the larger capacity of the neural network comes at the cost of slower predictions. For kernel methods situation is the same. The functional form generated by such methods is given by:
\begin{equation}
\text{prediction} = \sum\limits_q^{N_{samples}} c_q K(\text{train sample}_q, \text{test sample}),
\end{equation}
where the summation is over the whole training dataset or over the selected sparse points. Here again, the better fitting flexibility, which is determined by $ N_{samples}$, comes at the cost of a more considerable computational cost. For the linear models, there is the same tradeoff. For instance ACE\cite{drautz2019atomic}, MTP\cite{shapeev2016moment} and aPIP\cite{van2020regularised} express the energy as:
\begin{equation}
\label{eq:linear_basis}
\text{prediction} = \sum\limits_q^{N_{basis}} c_q B_q(\{\vec{r_i}\}),
\end{equation}
where $B_q(\{\vec{r_i}\})$ are the systematic basis functions of a collection of coordinates that describes the system. Here $N_{basis}$ plays the same role as the $N_{samples}$ for kernel methods. 

The fundamental feature of our potential, which also can be cast to the form of Eq. \ref{eq:linear_basis} is that the domain of the functions $B_q(\{\vec{r_i}\})$, where they are not zero, is finite, and thus, it is not necessary to evaluate all of them given a single chemical configuration. Even more, the number of basis functions to be evaluated stays constant and doesn't depend on the total number of the $N_{basis}$ basis functions used. The idea of finite support is also used in Polynomial Symmetry Functions (PSF) \cite{bircher2021improved} and in Ultra Fast (UF) potentials \cite{xie2021ultra}. In the case of PSF, it helps to significantly accelerate the computation of Behler-Parrinello symmetry functions\cite{behler2007generalized}, but later, on top of them, a neural network is applied, which shifts the overall computational cost from the conventional empirical potentials to the machine learning ones. Similar to GTTP, UF expresses two- and three-body potential in terms of the B-spline basis functions discussed above, but it lacks a number of important features. While we note that our potential can be cast to the form of eq. \ref{eq:linear_basis} in order to highlight the ultimate source of computational efficiency, in practice, the computational scheme of GTTP is more efficient than that. Once potential is fitted, we never evaluate the B-spline basis functions. Instead, we explicitly convert the resulting functional form to the spline parametrization (the conversion of the coefficients  $c_m$ and $d_{m_1, m_2, m_3}$ to $a_p^l$ and $b_{p_1, p_2, p_3} ^{l_1, l_2, l_3}$ mentioned in the previous section). Counting the required number of multiplications shows that this approach is way more efficient. For the two-body potential, in the case of the explicit evaluation of the basis functions, one needs to do $O(k^2)$ (where $k$ is the order of B-splines, practically we always use $k = 2$ for all the numerical experiments) multiplications per each pair of atoms. This number arises from the necessity to compute $k + 1$ basis functions, where each of them is given by a polynomial of order $k$. After the conversion to the spline parametrization, one needs to compute just one polynomial of order $k$, which costs only $O(k)$ multiplications. For the case of three-body potential, the difference is even more pronounced, $O(k^6)$ against $O(k^3)$ multiplications. On top of that, since we use symmetrization introduced in the eq.\ref{eq:symmetrization} it is possible to compute only one 3-dimensional polynomial for each \textbf{ordered} triplet of the atoms with the same specie in the system. 

We should note that our functional form is limited to two- and three-body interactions, and thus, our potential is not a universal approximator, which is also called systematically improvable, in contrast to the methods\cite{drautz2019atomic, van2020regularised, shapeev2016moment} discussed above. Though, as it will be shown later, for many systems, the possibility to approximate just two- and three-body potential arbitrarily close is already enough to achieve good accuracy.


Thus, the presented potential is in the speed group of conventional empirical potentials and at the same time is flexible enough to approximate arbitrary two- and three-body interactions without any additional assumptions.

\section{\label{sec:descriptors}New class of Invariant descriptors}
Usually, machine learning potentials are constructed in two steps. In the first step, a certain set of invariant descriptors is calculated, and in the second, it is fed to some machine learning algorithm. This is done because PES approximation should be invariant with respect to rotation, movement, reflection, and permutation of the identical atoms in the input structure. A good review of such descriptors is given in \cite{bartok2013representing}. It is clear that descriptors $D_{IJ_m}^2$ and $D_{IJK_{m_1, m_2, m_3}}^3$ satisfy all the mentioned requirements along with smoothness with respect to atomic coordinates and, therefore, can be used along with arbitrary smooth machine learning algorithms (e.g., neural networks and kernel methods with smooth kernel generate smooth functions, whereas some machine learning methods - e.g., random forest - do not). Atomic versions of these descriptors are meant to describe local atomic neighborhoods and are defined as:
\begin{widetext}
\begin{equation}
\overset{\text{atomic}}{D^2_{I_m}} = \sum\limits_{i \in \overset{\text{atomic}}{P_{I}}(R^2_{\text{cut}})} B_m^k(\mid\vec{r}_i -  \vec{r}_{\text{central}}\mid)
\end{equation}
and 

\begin{equation}
\overset{\text{atomic}}{{D^3_{IJ}}}_{m_1,m_2,m_3} = \sum\limits_{<i, j> \in \overset{\text{atomic}}{T_{IJ}}(R_{\text{cut}}^3)} \overset{\text{atomic}}{BS^k_{IJ}}_{m_1,m_2, m_3}(\mid\vec{r}_i -  \vec{r}_j\mid, \mid\vec{r}_i -  \vec{r}_{\text{central}}\mid, \mid\vec{r}_j -  \vec{r}_{\text{central}}\mid) \text{,}
\end{equation}
\end{widetext}
where $\overset{\text{atomic}}{P_{I}}(R_{cut})$ is the set of neighbors with type I, $ \overset{\text{atomic}}{T_{IJ}}(R_{\text{cut}}^3)$ is, analogously, the set of pairs of neighbors with types I and J and $\overset{\text{atomic}}{BS^k_{IJ}}_{m_1,m_2, m_3}$ are symmetric combinations of three-dimensional B-splines where the central atom is considered to be inequivalent to any of its neighbors regardless of its type. We leave the analysis of these descriptors and the relationships between our descriptors and Behler--Parinello symmetry functions \cite{behler2011atom} to future work.

\section{\label{sec:more_general}More general parametrization}
In the case of $\varphi_2(r)$, when piecewise polynomial parametrization with polynomials of order $k$ is used, there are $(k + 1) Q_2$ initial degrees of freedom. If the potential should be $k - 1$ times continuously differentiable, there are $k$ stitching conditions in all inner vertices of the grid and in the right outer vertice, $k Q_2$ in total. So, there are $(k + 1) Q_2 - k Q_2 = Q_2$ eventual degrees of freedom which corresponds to the $Q_2$ coefficients in the cardinal B-splines parametrization (in the form of Eq.~\ref{eq:new_parametrization}). But one can let the polynomials be of order $k$ and require the potential to be only $k_d - 1$ times continuously differentiable, where $k_d <  k$. In this case, there are $(k + 1 - k_d) Q_2$ eventual degrees of freedom. The corresponding cardinal B-splines parametrization is given by:

\begin{equation} \label{eq:new_general_parametrization}
    \varphi_2(r) = \sum_{m = 0}^{Q_2 - 1} \sum_{f = k_d}^{k} c_{f, m} B_m^f(r) \text{.}
\end{equation}

In the case of $\varphi_3$ three-dimensional cardinal B-splines of not uniform order are defined as:

\begin{equation}
    B_{m_1, m_2, m_3}^{f_1, f_2, f_3}(r_1, r_2, r_3) = B_{m_1}^{f_1}(r_1) B_{m_2}^{f_2}(r_2) B_{m_3}^{f_3}(r_3) \text{.}
\end{equation}
The definition of the symmetric combinations $BS^{f_1, f_2, f_3}_{m_1, m_2, m_3}$ is analogous to the Eq.~\ref{eq:symmetrization}, where in summation $f_1$, $f_2$ and $f_3$ are also rearranged along with $m_1$, $m_2$ and $m_3$. 
All subsequent steps including the definition of atomic invariant descriptors $\overset{\text{atomic}}{D^2_{I_{f, m}}} $ and $\overset{\text{atomic}}{{D^3_{IJ}}}_{f_1, f_2, f_3, m_1,m_2,m_3}$ are the same as before. 

When the training dataset is large enough, there is no need to use $k > k_d$. Indeed, one can just put $k = k_d$, not affecting the smoothness of the potential, and increase $Q_2$ and $Q_3$ to ensure the same fitting flexibility. After this procedure, the smoothness and accuracy of the potential will be the same as before, and computational time will be lower since polynomials of lower order will have to be calculated. 

But when the training dataset is not big enough, the use of $k > k_d$ may increase the accuracy of the potential since parametrization in the form of Eq.~\ref{eq:new_general_parametrization} along with lower $Q_2$ and $Q_3$ or bigger grid \text{intervals} may have better generalization capability. 

\section{\label{sec:results}results}

\subsection{Aluminum}
Aluminum is an example of a system where two- and three-body interaction approximation works well.
To illustrate the performance of our potential, we applied it to four datasets. The first one contains 5000 steps of \textit{ab initio} molecular dynamics simulation in the canonical ($NVT$) ensemble of aluminum with 108 atoms in the unit cell at 300 K and with volume $16.7 \frac{\si{\angstrom}^3}{\text{atom}}$. The second dataset consists of 20000 random structures produced by a symmetric random structure generator from evolutionary algorithm USPEX \cite{oganov2006crystal, oganov2011evolutionary, lyakhov2013new}, each with 8 atoms, third is a subset of the second one and contains 7071 structures with negative energies and fourth is a subset of the third one and contains 2088 structures with energies less than $-3.13\: \frac{eV}{atom}$. The overview of these datasets is given in Table \ref{table:Al_datasets}. All \textit{ab initio} calculations of energies and forces were performed using Vienna Ab initio Simulation Package (VASP) \cite{kresse1993ab, kresse1996efficient, kresse1996efficiency}. Projector-augmented wave (PAW)\cite{kresse1999ultrasoft} method was used to describe core electrons and their interaction with valence electrons. The plane wave kinetic energy cutoff was set at $500\:eV$ and $\Gamma$-centered k-points with a resolution of $2\pi\times0.05\:\si{\angstrom}^{-1}$ were used.

\begin{table*}
\centering
\caption {Summary of aluminum datasets. $N_s$ means the  number of structures, $N_a$ is the number of atoms in unit cell. In this particular case, it is identical for all structures within one dataset. $F$ means scalar force components---projections on x, y and z axes.} \label{table:Al_datasets}
    \begin{tabular}{ | l | l | l | | l | |l| |l| |l| |l| }    
    \hline
    Notation & $N_s$ & $N_a$ & min $E$, $\frac{eV}{Atom}$ & max $E$, $\frac{eV}{Atom}$ & $\overline{E}$, $\frac{eV}{Atom}$  &  $\sqrt{\overline{(E - \overline{E})^2}} $, $\frac{eV}{Atom}$ & $\sqrt{\overline{F^2}}$, $\frac{eV}{\si{\angstrom}}$ \\ \hline
    $\text{Rand}_1$ & 20000 & 8 & -3.75 & 54.84 & 4.06 & 7.15 & 22.87 \\ \hline 
$\text{Rand}_2$ & 7071 & 8 & -3.75 & -0.00 & -2.15 & 1.15 & 5.00 \\ \hline 
$\text{Rand}_3$ & 2088 & 8 & -3.75 & -3.13 & -3.48 & 0.19 & 0.69 \\ \hline 
MD & 5000 & 108 & -3.75 & -3.69 & -3.71 & 0.0034 & 0.35 \\ \hline 

    \hline
    \end{tabular}
\end{table*}

The following several subsubsections contain an analysis of the hyperparameters of the developed potential. For the sake of brevity, thereinafter, we will understand forces as force components---projections on the x, y, and z axes. Error in energies per atom is a rather unphysical quantity since the total error per unit cell does not necessarily grow proportionally to the number of atoms in it. So, we decided to give all errors in energies per unit cell. Relative errors are calculated as the ratio of the absolute errors to the standard deviations of the corresponding values. All errors are given on the test samples and were obtained either by cross-validation or by explicit partitioning into train and test sets. 
\subsubsection{Relative importance}
When solving the linear regression problem, the following minimization problem arises:
\begin{equation}
\begin{split}
    &\min_{c, d} \frac{1} {\lambda} (\sum_{i} c^2_i + \sum_i d^2_i) +\\& \frac{W_E}{N_E} \sum_{i} (E_{\textit{ab initio}_i} - E_{predicted_i}(c, d)) ^ 2 + \\& \frac{W_F}{N_F} \sum_{i} (F_{\textit{ab initio}_i} - F_{predicted_i}(c, d)) ^ 2 \text{,}
\end{split}
\end{equation} where $W_E$ and $W_F$ are weights for the energies and forces,  $N_E$ and $N_F$ are numbers of energies and forces in the dataset. $\lambda$ is the usual $L_2$ regularization hyperparameter, which can be selected using standard techniques \cite{friedman2001elements, mackay1992bayesian}, while the influence of $Im = W_E / W_F$ - relative importance of energies, should be investigated manually. 

\begin{figure}[h]
    \centering
    \includegraphics[width=8cm]{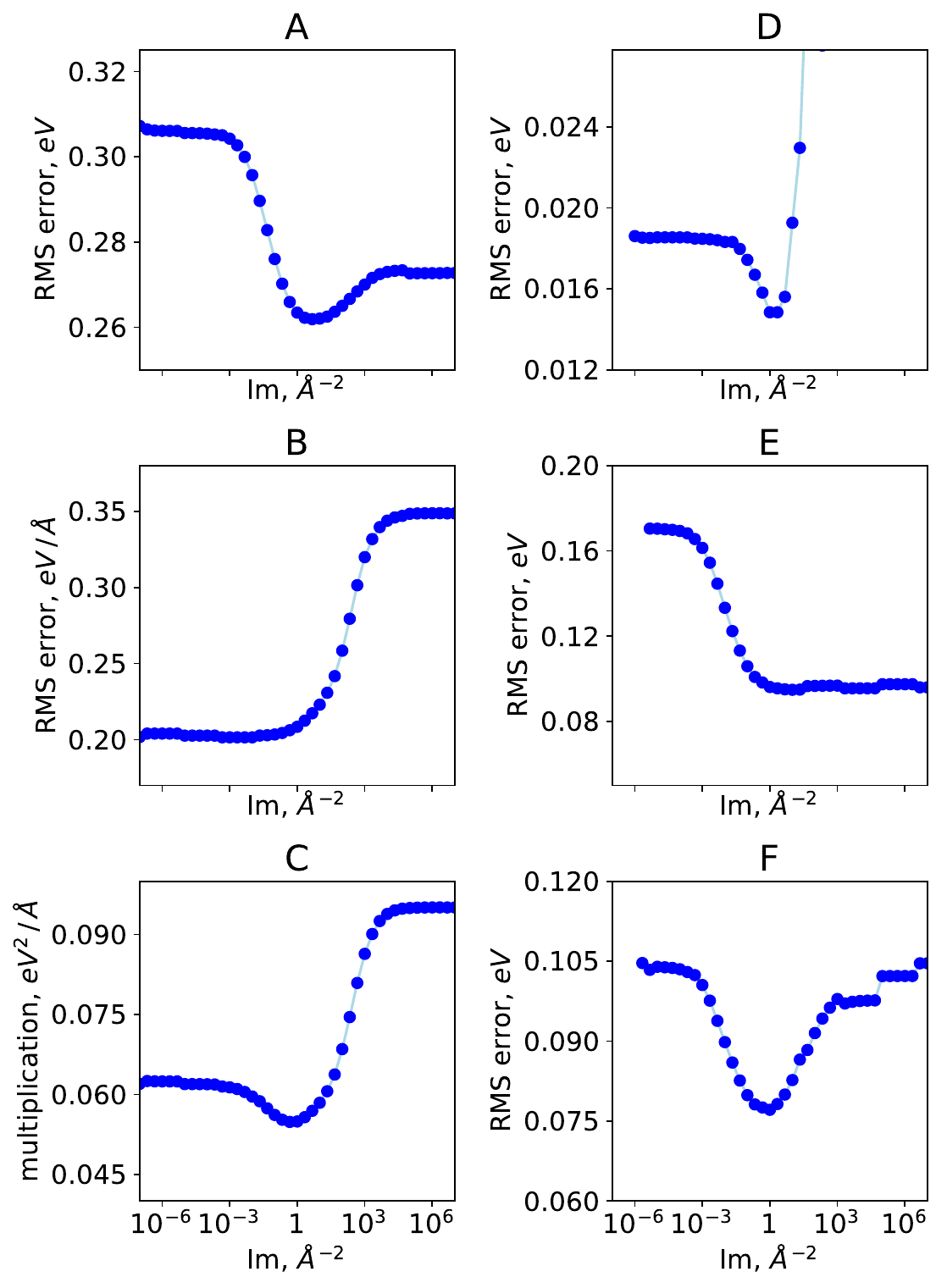}
    \caption{influence of the relative importance of energies, $Im$ hyperparameter. Panels A--C are related to $\text{Rand}_2$ dataset and illustrate cross-validation RMS errors in energies, forces, and their product respectively. Panels D--F illustrate errors in energies. D corresponds to the potential trained on one-tenth of the MD dataset, E and F to the potentials with a small and large number of parameters, respectively, trained on $\text{Rand}_3$ dataset.}
    \label{fig:importance_choosing}
\end{figure}

First of all, we investigated it on $\text{Rand}_2$ dataset. The other hyperparameters of the potential were put as $S_2 = S_3 = \SI{1.0}{\angstrom}$, $R^2_{\text{cut}} = \SI{10.0}{\angstrom}$, $R^3_{\text{cut}} = \SI{5.0}{\angstrom}$, $Q_2 = 27$, $Q_3 = 8$, $k = 2$, first variant of triples cutting. 
For each value of $Im$ we measured RMS error in energies and forces. All errors were evaluated by 20-fold cross-validation with random partitions. Results are shown in Fig. \ref{fig:importance_choosing}. 

It is very natural that the higher the value of $Im$, or, in other words, the higher priority the energies are given, the lower the error in energies and vice versa. But there is also another effect. The thing is that the number of energies in the dataset is much less than the number of forces. Indeed, structure, which contains $N_a$ atoms, contributes one energy and $3N_a$ forces to the dataset. Thus, energies alone typically do not provide enough data to fit the potential, and training only on energies leads to overfitting. When the value of $Im$ is very large, the potential is actually trained only on energies. So, one can expect that decreasing $Im$ or taking into account the forces during the fitting can reduce the test error in energies. Fig. \ref{fig:importance_choosing} A, D, E, and F illustrate the dependence of test error in energies on the $Im$ for different datasets and different potentials. In accordance with the reasons discussed earlier, all these dependencies consist of two plateaus and a well between them. The relative position of the plateaus and the size of the well depend on the interrelation between dataset size and the number of parameters in the potential.

Fig. \ref{fig:importance_choosing} B illustrates the errors in forces. We observe qualitatively similar behavior in all studied cases.

Since we assume that the errors in energies and forces are equally important, we decided to choose the value of $Im$ to minimize the product of these errors, which is plotted in Fig. \ref{fig:importance_choosing} C. 

\subsubsection{Two-body hyperparameters}

\begin{figure}[h]
    \centering
    \includegraphics[width=8cm]{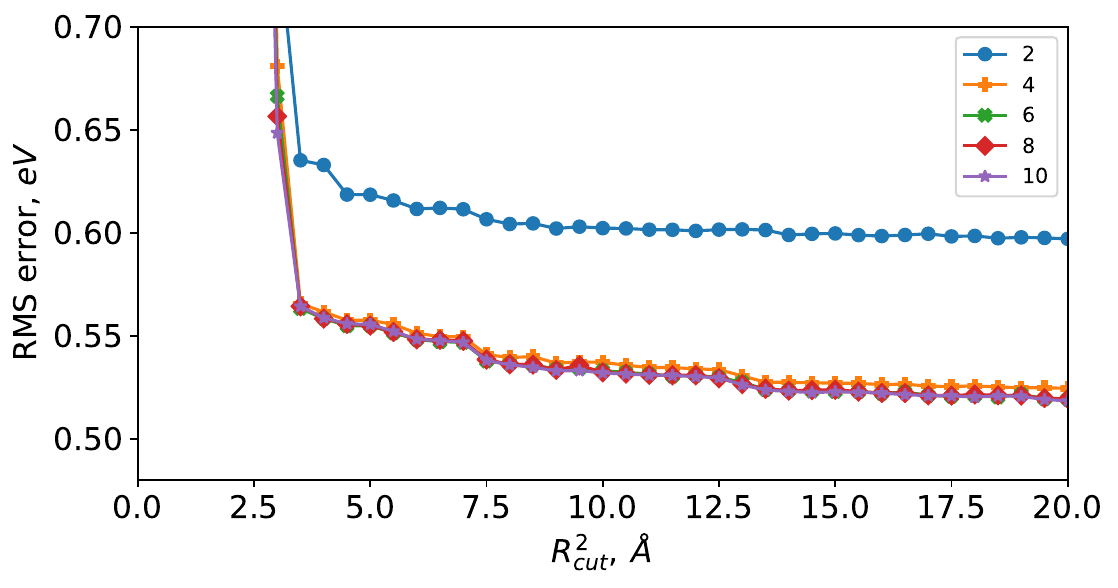}
    \caption{Cross-validation RMS errors in energies for only two-body potential.}
    \label{fig:two_particles_convergence}
\end{figure}

$R^2_{\text{cut}}$ and $R^3_{\text{cut}}$ represent the tradeoff between the accuracy and computational time. The higher $R^2_{\text{cut}}$, the more accurate the potential, but also slower. We measured the behavior of the error in energies for only two-body potential at various $R^2_{\text{cut}}$ and different grid densities, namely $2$, $4$, $6$, $8$ and $10$ $\text{\text{intervals}}/\si{\angstrom}$, on the $\text{Rand}_2$ dataset. Results are shown in Fig. \ref{fig:two_particles_convergence}. 

As can be seen from this plot, the RMS error converges to some non-zero limit, which is the limit of the accuracy of the two-body approximation. 

For later calculations we have chosen $R^2_{\text{cut}} = \SI{8}{\angstrom}$ and $Q_2$ corresponding to the grid density of $6$ $\text{intervals}/\si{\angstrom}$ as hyperparameters at which the error almost completely converged. 
\subsubsection{Three-body hyperparameters}

Now we fix hyperparameters of two-body potential found previously and measure the performance of two- and three-body potential with different three-body hyperparameters. As earlier, we performed calculations for various $R^3_{\text{cut}}$ and several grid densities. 

\begin{figure}[h]
    \centering
    \includegraphics[width=8cm]{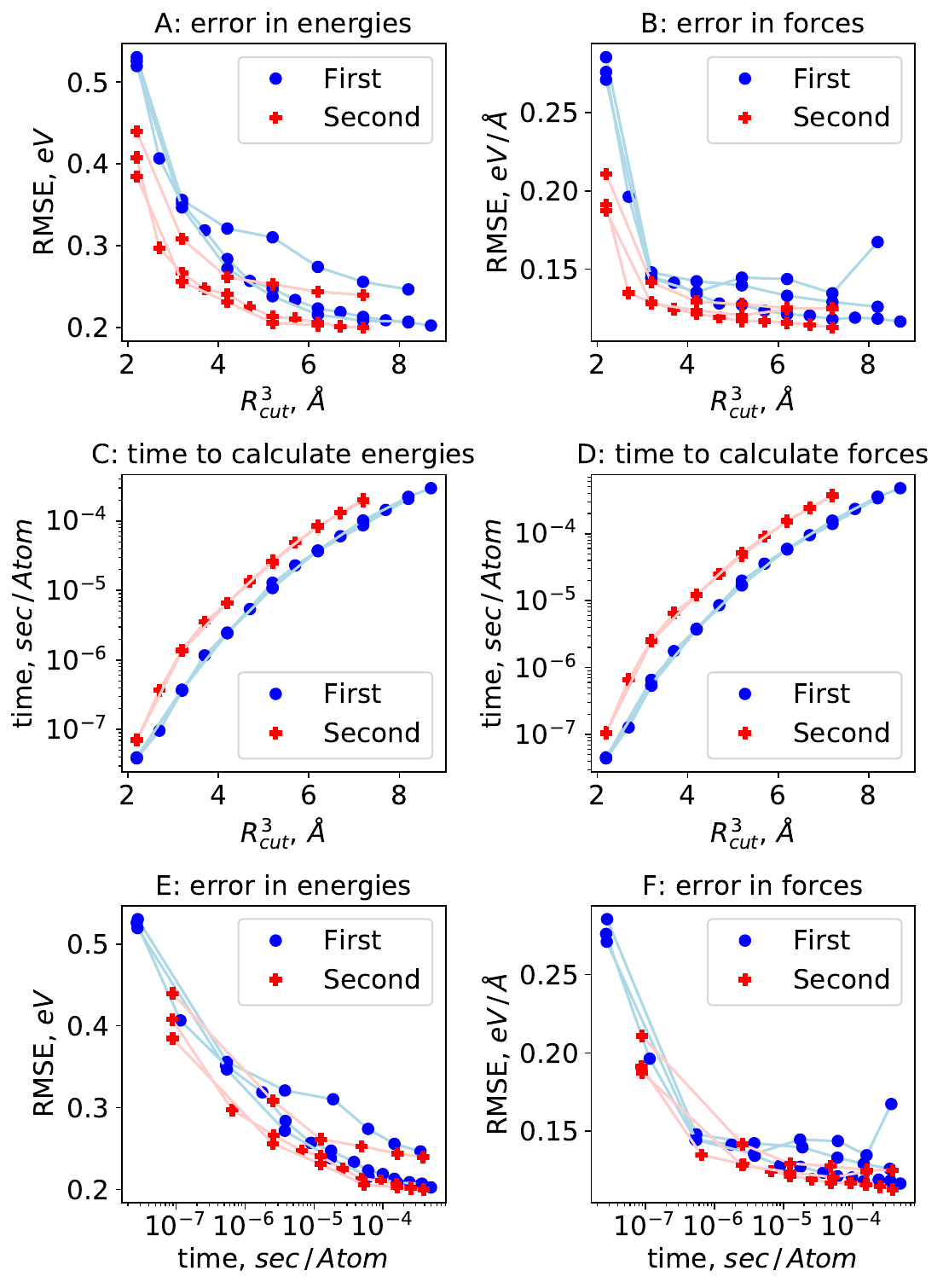}
    \caption{Influence of three-body hyperparameters. All panels contain lines for several grid densities, namely $1$, $2$, and $3$ $\text{intervals}/\si{\angstrom}$, and first and second variants of triples cutting. Subplots A and B illustrate errors in energies and forces for different $R_{\text{cut}}^3$,  C and D show computational time for energies and forces. E and F present a tradeoff between computational time and errors. Time on the horizontal axis corresponds to the simultaneous calculation of both energies and forces. All measurements were taken on one core of Intel(R) Xeon(R) CPU E5-2667 v4 for only three-body part, not including the construction of atomic neighborhoods. Times were averaged over a set of structures from the $\text{Rand}_2$ dataset. All standard errors of the mean do not exceed the size of the symbols.}
    \label{fig:three_particles_convergence}
\end{figure}

Fig. \ref{fig:three_particles_convergence} A and \ref{fig:three_particles_convergence} B illustrate the behavior of errors in energies and forces with increasing $R^3_{\text{cut}}$. As expected at the same $R^3_{\text{cut}}$, the error is lower with the second variant of triples cutting because at the same $R^3_{\text{cut}}$ the set of considered triples with the first variant of triples cutting is a subset of triples included with the second variant of triples cutting. But, for the same reason, the computational time with the second variant of triples cutting is higher at the same $R^3_{\text{cut}}$, as illustrated in Fig.  \ref{fig:three_particles_convergence} C and \ref{fig:three_particles_convergence} D. These figures also show that computational time indeed does not depend on $Q_2$ and $Q_3$ (lines for different densities almost coincide), and thus on fitting flexibility. 

Fig. \ref{fig:three_particles_convergence} E and \ref{fig:three_particles_convergence} F illustrate the tradeoff between accuracy and computational time. It appears that for this particular chemical system, the
the second variant of triples cutting is slightly better.

We consider the $R^3_{\text{cut}} = \SI{5.2}{\angstrom}$ with second variant of triples cutting as sufficient. $Q_3$ was chosen to correspond grid density equal to $3$ $\text{intervals}/\si{\angstrom}$. 

\begin{figure}[h]
    \centering
    \includegraphics[width=8cm]{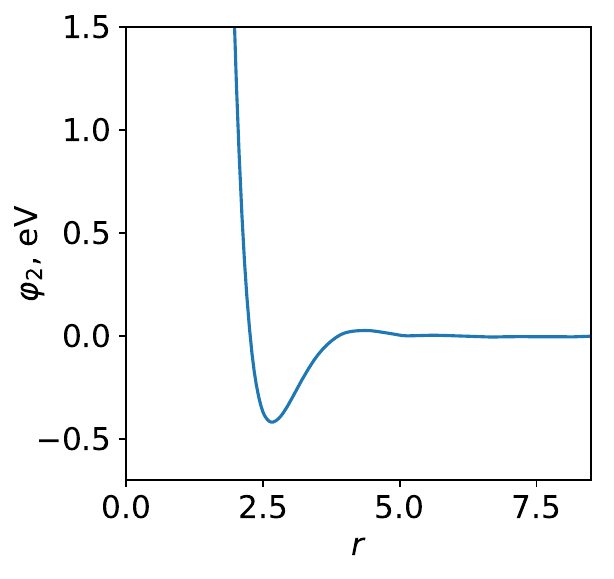}
    \caption{Two-body potential trained on $\text{Rand}_2$ dataset for Al. }
    \label{fig:2_body_potential_Al}
\end{figure}

\begin{figure}[h]
    \centering
    \includegraphics[width=8cm]{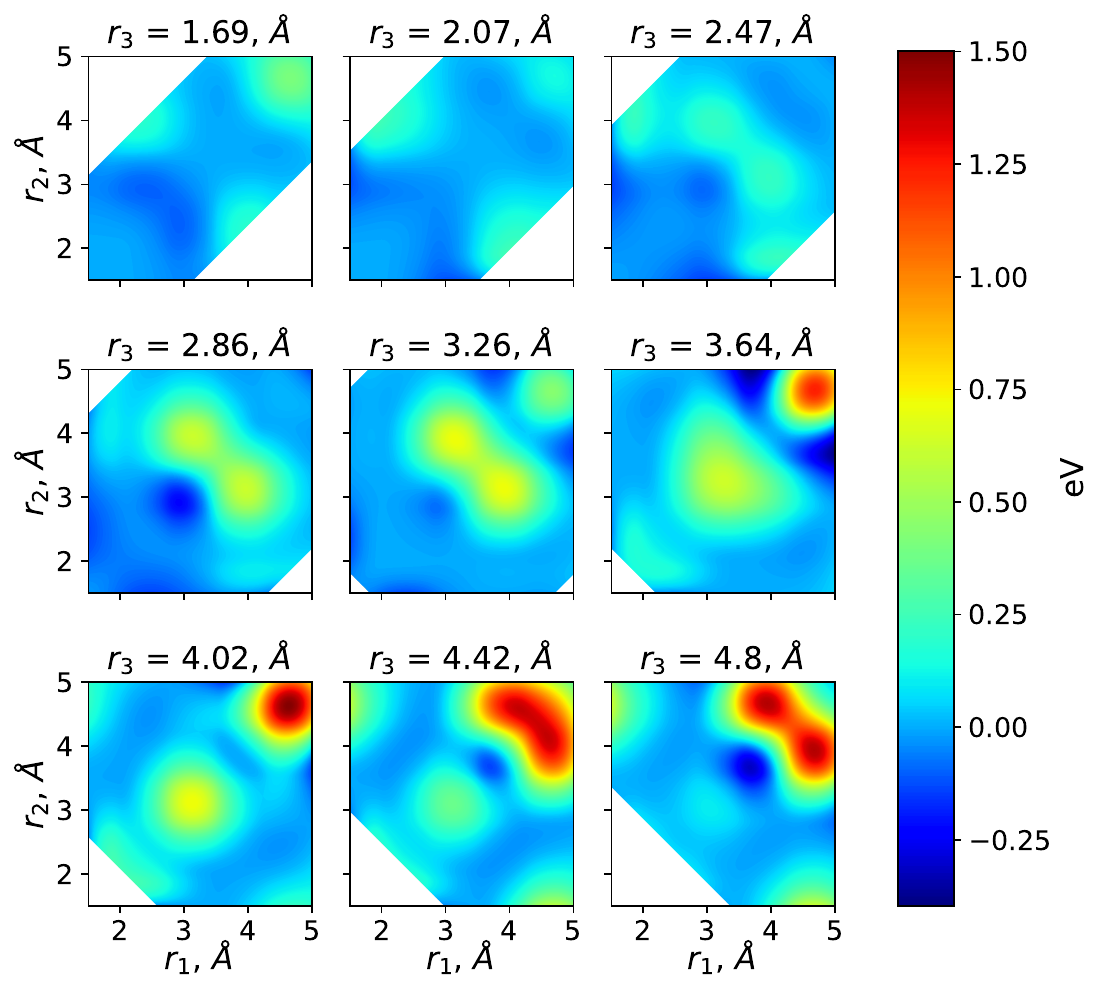}
    \caption{Three-body potential trained on $\text{Rand}_2$ dataset for Al. }
    \label{fig:3_body_potential_Al}
\end{figure}

The resulting two- and three-body potentials are shown in Fig. \ref{fig:2_body_potential_Al} and \ref{fig:3_body_potential_Al}. We independently calculated two- and three-body contributions to the energy, and it appeared that the three-body part is an order of magnitude smaller. Namely, standard deviations of two- and three-body components on the $\text{Rand}_2$ dataset appeared to be $9.47\:eV$ and $1.30\:eV$,  respectively.

\subsubsection{Performance summary}
\label{sec:al_performance}

\begin{figure}[h]
    \centering
    \includegraphics[width=8cm]{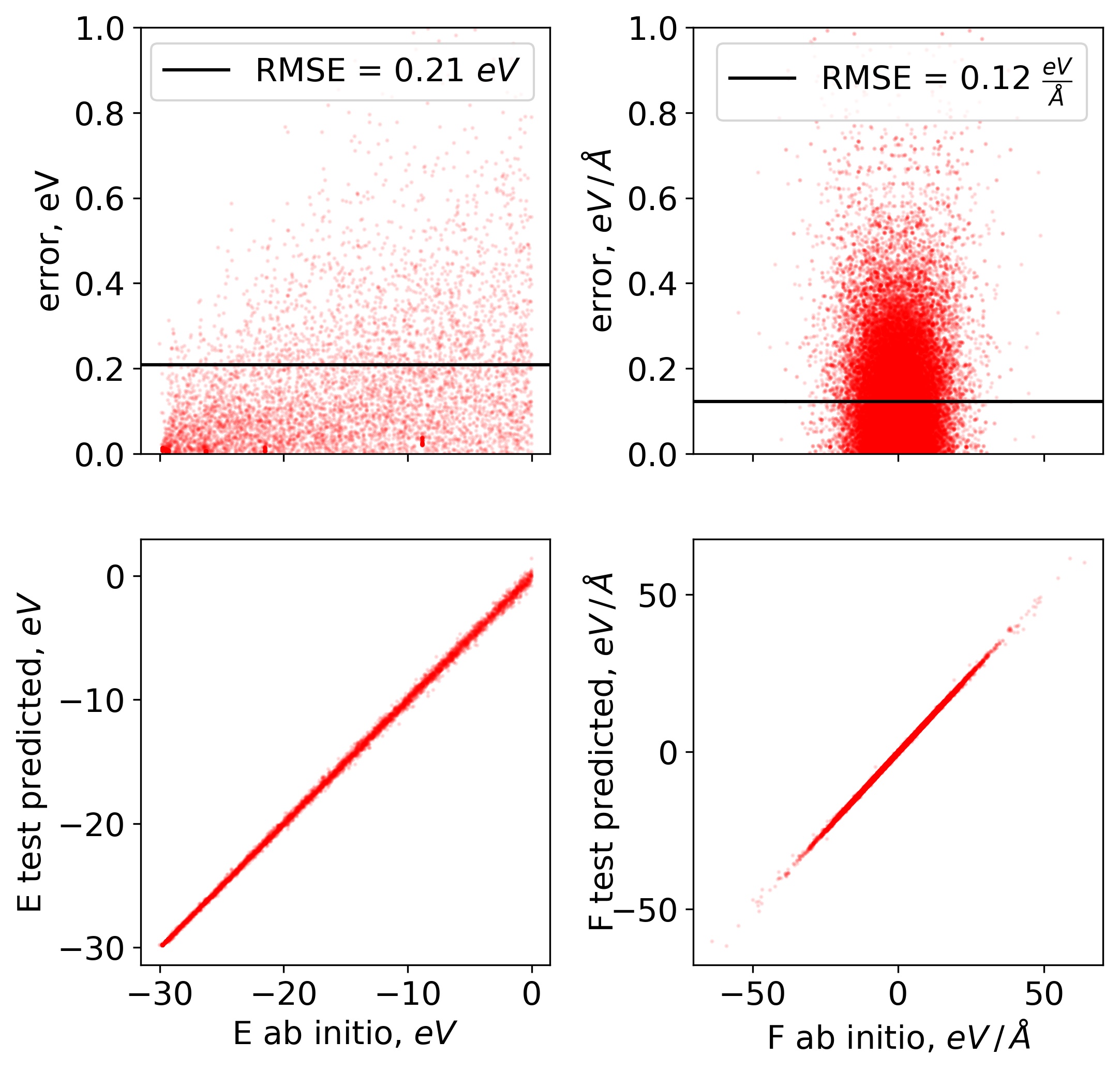}
    \caption{Performance of the GTTP on the  $\text{Rand}_2$ dataset. Energies and forces are predicted in the cross-validation cycle on the test samples.}
    \label{fig:Al_intermediate_performance}
\end{figure}

Performance of the potential on the $\text{Rand}_2$ dataset is illustrated in  Fig. \ref{fig:Al_intermediate_performance}.

\begin{table*}
\caption {Performance of GTTP for Al on energies. Absolute RMS errors are given in $meV$ per unit cell(8 atoms/cell in case of $\text{Rand}_{\alpha}$ and 108 atoms/cell in case of MD). Relative errors are calculated as the ratio of the absolute error to the standard deviation.  $\text{Rand}_{\alpha}$ -- MD cells illustrates errors after additive constant adjusting. See Fig. \ref{fig:eq_md_performance} and discussion in the text.} \label{table:performance_on_energies}
    \begin{tabular} {| C{2.2cm} | C{2cm} | C{2cm} | C{2cm} | C{2cm} | C{2cm} |}
    \hline
    \diagbox[width=2.2cm]{train on}{test on} & \multicolumn{2}{c|}{MD}  & $\text{Rand}_3$ & $\text{Rand}_2$ & $\text{Rand}_1$\\ 
    \hline
 MD & 7.4, 2.01\% & 9.3, 2.55\% & \crossedout & \crossedout & \crossedout \\
\hline
$\text{Rand}_3$ & \multicolumn{2}{c|}{42.9, 11.74\%} & 60.6, 3.89\% & \crossedout & \crossedout \\
\hline
$\text{Rand}_2$ & \multicolumn{2}{c|}{119.1, 32.61\%}  & 71.2, 4.56\% & 208.0, 2.27\% & \crossedout \\
\hline
$\text{Rand}_1$ & \multicolumn{2}{c|}{111.7, 30.57\%}  & 202.6, 12.99\% & 393.8, 4.3\% & 990.2, 1.73\% \\
\hline
    \end{tabular}
\end{table*}

\begin{table*}
\caption {Performance of GTTP for Al on forces. Absolute RMS errors are given in $\frac{meV}{\si{\angstrom}}$. Relative errors are calculated as the ratio of the absolute error to the standard deviation.} \label{table:performance_on_forces}
    \begin{tabular}{| C{2.2cm} | C{2cm} | C{2cm} | C{2cm} | C{2cm} | C{2cm} |}
    \hline
    \diagbox[width=2.2cm]{train on}{test on} & \multicolumn{2}{c|}{MD} & $\text{Rand}_3$ & $\text{Rand}_2$ & $\text{Rand}_1$\\ 
    \hline
 MD & 12.0, 3.47\% & 12.3, 3.55\% & \crossedout & \crossedout & \crossedout \\
\hline
$\text{Rand}_3$ & \multicolumn{2}{c|}{41.2, 11.85\%} & 27.7, 4.0\% & \crossedout & \crossedout \\
\hline
$\text{Rand}_2$ & \multicolumn{2}{c|}{86.6, 24.94\% } & 34.6, 5.01\% & 121.8, 2.44\% & \crossedout \\
\hline
$\text{Rand}_1$ & \multicolumn{2}{c|}{75.1, 21.63\%}  & 58.9, 8.52\% & 157.6, 3.15\% & 625.6, 2.74\% \\
\hline
    \end{tabular}
\end{table*}

For the other datasets, optimal hyperparameters of the potential were chosen in a similar manner, and they do not differ much.

The numerical overview is given in Tables \ref{table:performance_on_energies} and \ref{table:performance_on_forces}.
Dataset $\text{Rand}_3$ is a subset of $\text{Rand}_2$, which in turn is a subset of $\text{Rand}_1$. In $\text{Rand}_\alpha$--$\text{Rand}_\beta$ cells all energies and forces are predicted in a cross-validation cycle for $\text{Rand}_\alpha$ dataset with hyperparameters of the potential selected for $\text{Rand}_\alpha$, and later the error is measured only on values, which belong to $\text{Rand}_\beta$.

$\text{Rand}_{\alpha}$ -  MD cells illustrate the errors on MD of the potentials trained on $\text{Rand}_{\alpha}$. In the case of energies, these cells illustrate the errors after additive constant adjusting. Indeed, initially, there is a constant systematic error, see Fig. \ref{fig:eq_md_performance}. It originates from both discrepancy between \textit{ab initio} calculations and intrinsic error of the potential. In the case of different datasets, namely $\text{Rand}$ and MD, \textit{ab initio} calculations were performed with different parameters, which led to different ground state energies in both cases. Also, the potential itself predicts the ground state energy is not absolutely correct. While contributing a relatively small part to the $\text{Rand}_\alpha$--$\text{Rand}_\beta$ errors, this makes a noticeable contribution in the case of $\text{Rand}_\alpha$--MD because the variability in the $\text{Rand}_{\alpha}$ datasets is much greater than in the MD, see Table \ref{table:Al_datasets}.  

The left subcell of MD--MD in Table \ref{table:performance_on_energies} illustrates the "interpolation" error when the error is measured in a cross-validation cycle with random partitions, while the right subcell illustrates the "extrapolation" error when potential is trained on the first third of the timeline of molecular dynamics and tested on the last.

Thus, all errors presented in Tables \ref{table:performance_on_energies} and \ref{table:performance_on_forces} are measured on test samples. 

Generally, the absolute error significantly depends on the variability in the dataset. The smaller part of phase volume is covered by the potential---the smaller is the absolute error and vice versa. Tables \ref{table:performance_on_energies} and \ref{table:performance_on_forces} also illustrate good transferability of the potential---being fitted to the beginning of the molecular dynamics trajectory, it can accurately describe system states from the last MD steps. In addition, it can, with satisfactory accuracy, predict energies and forces for structures with 108 atoms, being fitted to only structures with 8 atoms. Taking into account that the computational cost of acceptable accurate \textit{ab initio} calculations scales cubically with system size, this property is especially useful. The performance on the MD dataset of the potential trained on $\text{Rand}_3$ is shown in Fig.  \ref{fig:eq_md_performance}:

\begin{figure}[h]
    \centering
    \includegraphics[width=8cm]{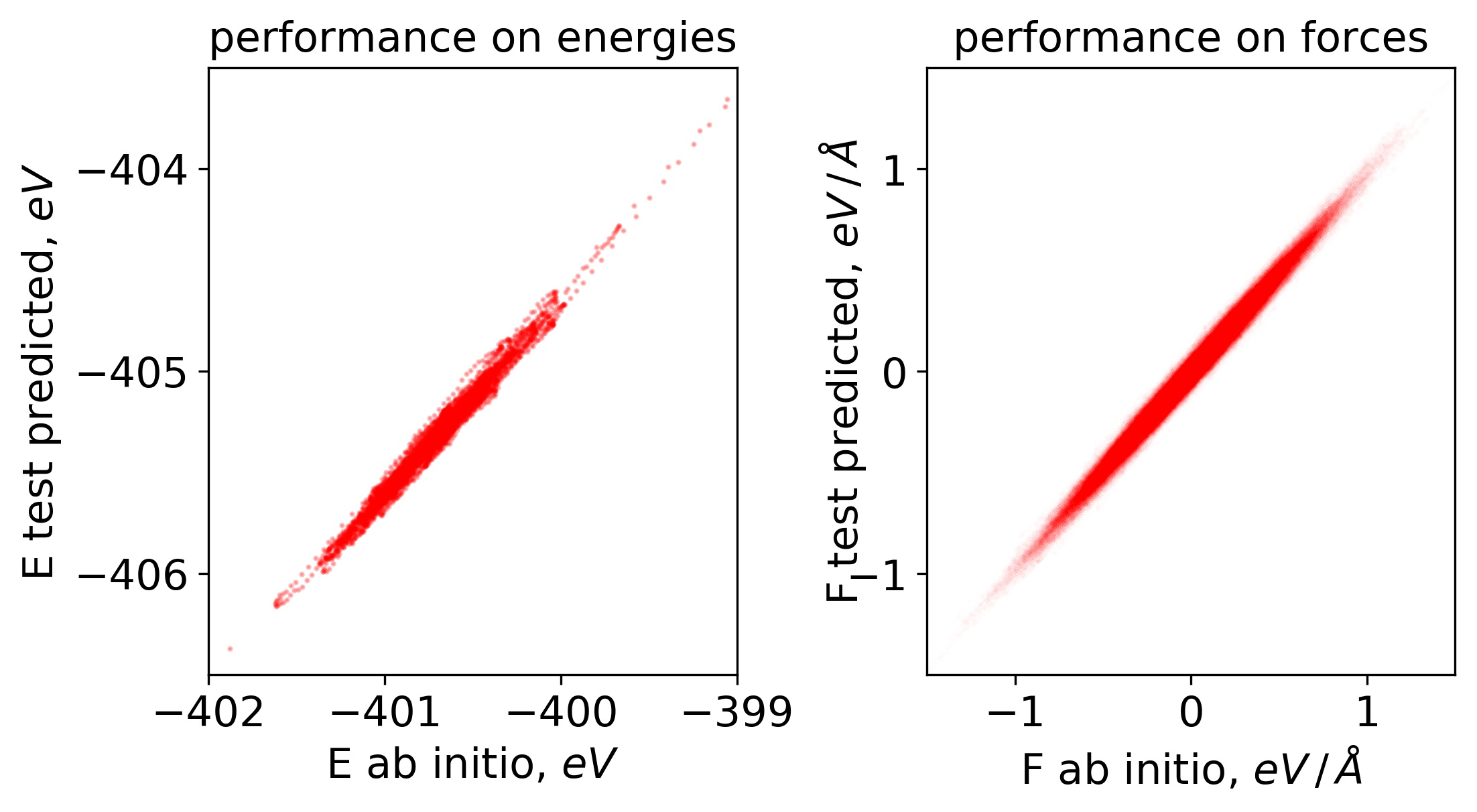}
    \caption{Performance on the MD dataset of the potential trained on $\text{Rand}_3$ before additive constant adjusting. Note the discrepancy between vertical and horizontal axes in the energy graph, as discussed in the text.}
    \label{fig:eq_md_performance}
\end{figure}

\subsubsection{Computational time}

\begin{figure}[h]
    \centering
    \includegraphics[width=8cm]{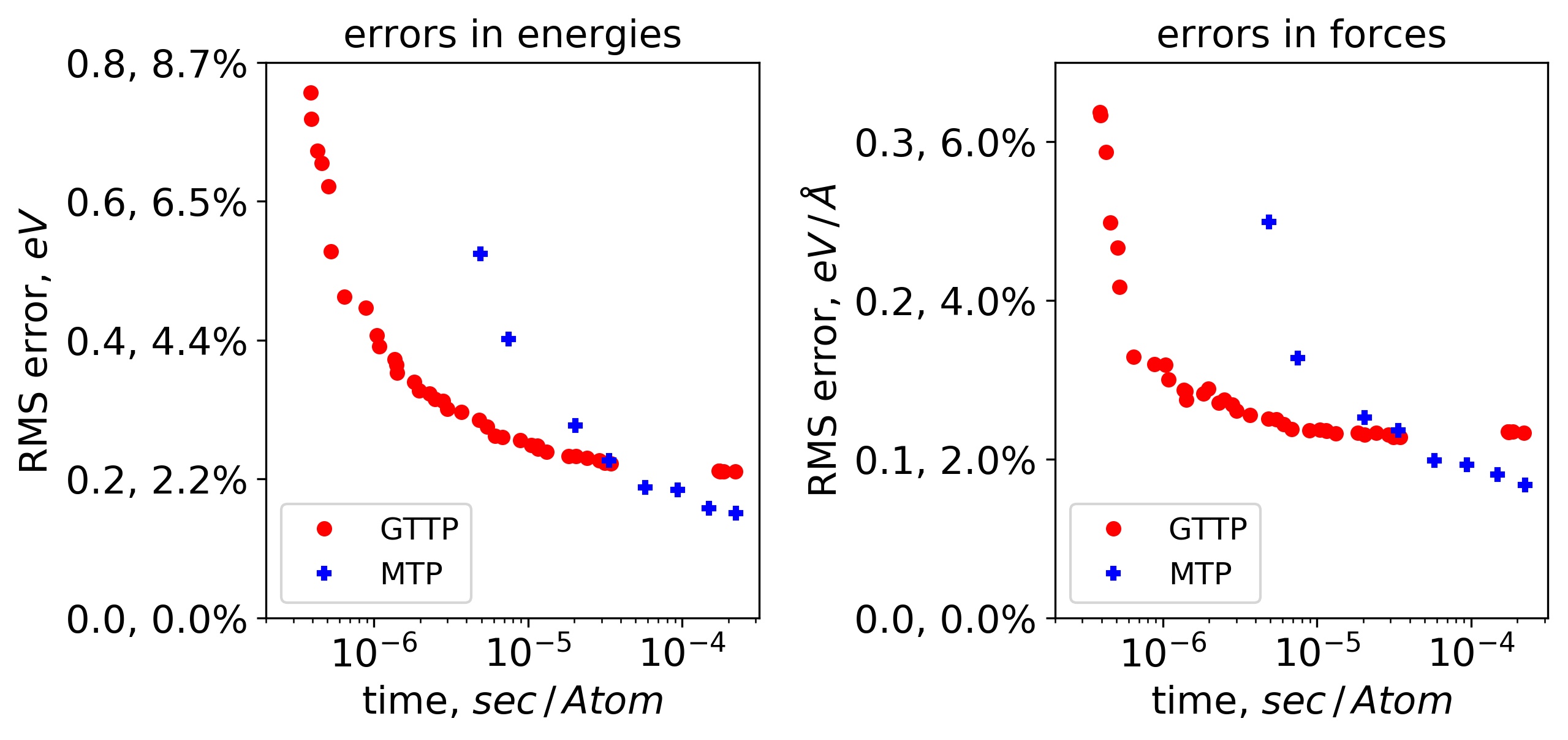}
    \caption{Accuracy--computational time tradeoff. Times were averaged over a set of structures from the $\text{Rand}_2$ dataset. All standard errors of the mean do not exceed the size of the symbols.}
    \label{fig:pareto}
\end{figure}

The hyperparameters of the potentials in previous sections were chosen far in saturation area, while it is possible to take smaller $R^2_{\text{cut}}$ and $R^3_{\text{cut}}$ to significantly reduce computational time and only slightly affect the accuracy. In order to investigate the tradeoff between time and accuracy, we fitted a number of potentials with different two- and three-body hyperparameters on the $\text{Rand}_2$ dataset. After that, we constructed the two-objective Pareto front, the first objective being computational time and the second one being the product of errors in energies and forces. To estimate errors, we used explicit partitioning into the train and test dataset with 80\% of the structures in the training dataset. Times were measured within LAMMPS Molecular Dynamics Simulator\cite{plimpton1995fast} to simultaneously calculate energies, forces, and stress tensors, including constructing atomic neighborhoods on one core of Intel(R) Xeon(R) CPU E5-2667 v4. Also, we compared the Pareto front of our (GTTP) potential with the Pareto front of the Moment Tensor Potential (MTP)\cite{ShapeevPareto}. The method of measuring time was the same in both cases. The result is shown in Fig.  \ref{fig:pareto}.

\begin{table*}
\caption{Accuracy--computational time tradeoff on the $\text{Rand}_2$ dataset. Columns time $E$ and time $F$ are related to the separate calculation of energies and forces outside the LAMMPS Molecular Dynamics Simulator not including construction of atomic neighborhoods. Total time represents the time to simultaneously calculate energies, forces, and stress tensors within the LAMMPS Molecular Dynamics Simulator, including the construction of atomic neighborhoods. The total time should not necessarily be more than the sum of the times to calculate energies and forces separately because these calculations share a large amount of work, such as, for example, calculation of monomials $r_1^{l_1} r_2^{l_2} r_3^{l_3}$. Times were averaged over a set of structures from the $\text{Rand}_2$ dataset. The given errors represent standard errors of the mean.} 
     \begin{tabular}{| >{\centering} m{2.1cm} |  >{\centering} m{2.1cm} | >{\centering}m{2.4cm} | >{\centering}m{1.1cm} | >{\centering}m{1.1cm} | >{\centering}m{1.1cm} | >{\centering}m{2.2cm} | >{\centering}m{1.9cm} | @{}m{0cm}@{}}
    \hline
    time E, $\frac{\mu s}{atom}$ & time F, $\frac{\mu s}{atom}$ & total time, $\frac{\mu s}{atom}$ & $R^2_{\text{cut}}$, $\si{\angstrom}$ & $R^3_{\text{cut}}$, $\si{\angstrom}$ & triples cutting &  RMSE E, $eV$ & RMSE F, $\frac{eV}{\si{\angstrom}}$ & \\ [0.6em] \hline
    
0.031 $\pm$ 0.001 & 0.048 $\pm$ 0.002 & 0.29 $\pm$ 0.01 & 2.5 &  -  & - & 1.16, 12.69\%  & 0.50, 9.98\%  & \\ \hline
0.046 $\pm$ 0.002 & 0.071 $\pm$ 0.003 & 0.39 $\pm$ 0.01 & 2.7 &  -  & - & 0.72, 7.84\%  & 0.32, 6.33\%  & \\ \hline 
0.118 $\pm$ 0.006 & 0.18 $\pm$ 0.01 & 0.51 $\pm$ 0.02 & 2.7 & 2.2 & second & 0.62, 6.78\%  & 0.23, 4.66\%  & \\ \hline  
0.17 $\pm$ 0.007 & 0.24 $\pm$ 0.01 & 0.64 $\pm$ 0.02 & 2.8 & 2.8 & first & 0.46, 5.04\%  & 0.16, 3.29\%  & \\ \hline 
0.51 $\pm$ 0.02 & 0.92 $\pm$ 0.04 & 1.42 $\pm$ 0.06 & 2.8 & 2.8 & second & 0.35, 3.85\%  & 0.14, 2.75\%  & \\ \hline 
1.15 $\pm$ 0.05 & 2.12 $\pm$ 0.08 & 3.0 $\pm$ 0.1 & 3.1 & 3.1 & second & 0.30, 3.28\%  & 0.13, 2.61\%  & \\ \hline 
2.43 $\pm$ 0.09 & 4.4 $\pm$ 0.2 & 6.1 $\pm$ 0.2 & 3.5 & 3.5 & second & 0.26, 2.86\%  & 0.12, 2.44\%  & \\ \hline 
5.8 $\pm$ 0.2 & 11 $\pm$ 0.4 & 13.3 $\pm$ 0.5 & 4.0 & 4.0 & second & 0.24, 2.61\%  & 0.12, 2.32\%  & \\ \hline 
14.5 $\pm$ 0.3 & 25.3 $\pm$ 0.5 & 31.4 $\pm$ 0.5 & 4.7 & 4.7 & second & 0.22, 2.43\%  & 0.11, 2.28\%  & \\ \hline 
15.9 $\pm$ 0.3 & 27.6 $\pm$ 0.5 & 34.6 $\pm$ 0.7 & 4.8 & 4.8 & second & 0.22, 2.42\%  & 0.11, 2.27\%  & \\ \hline 
126 $\pm$ 2 & 200 $\pm$ 3 & 221 $\pm$ 2 & 10.2 & 6.2 & second & 0.21, 2.29\%  & 0.12, 2.33\%  & \\ \hline    
    \end{tabular}
\label{table:pareto}
\end{table*}

MTP is one of the fastest machine learning potentials. Namely, it was shown\cite{shapeev2016moment} that on the same dataset with the same accuracy, MTP is approximately 170 times faster than the Gaussian Approximation Potential (GAP)\cite{bartok2010gaussian}. This was also confirmed in a recent study \cite{zuo2019performance}, where a comprehensive comparison of several machine learning potentials was performed. In spite of this, our potential convincingly outperforms MTP in the fast area. With increasing the computational time, the error of the GTTP converges to a non-zero limit which is caused by the error of the two- and three-body interactions approximation itself. When this happens, the error of the systematically
improvable  MTP becomes lower. In case of forces the convergence is reached already at $10^{-6} \frac{\text{sec}}{\text{atom}}$, whereas in the case of energies it is reached at $10^{-6} - 10^{-5} \frac{\text{sec}}{\text{atom}}$. Some potentials from the Pareto front are shown in Table \ref{table:pareto} 

When there are more than one atomic species, the potential energy surface is more complex, and, therefore, more parameters are required. Particularly, in GTTP, the number of parameters grows cubically with the number of atomic species. But at the same time, the computational cost of our potential does \textbf{not} increase with the number of parameters or with the number of atomic species. This is not the case for the majority of machine learning potentials and of MTP in particular, so one can expect that the relative performance of our potential will be even better on multicomponent systems. 
   
\subsection{Tungsten}

\begin{figure}[h]
    \centering
    \includegraphics[width=8cm]{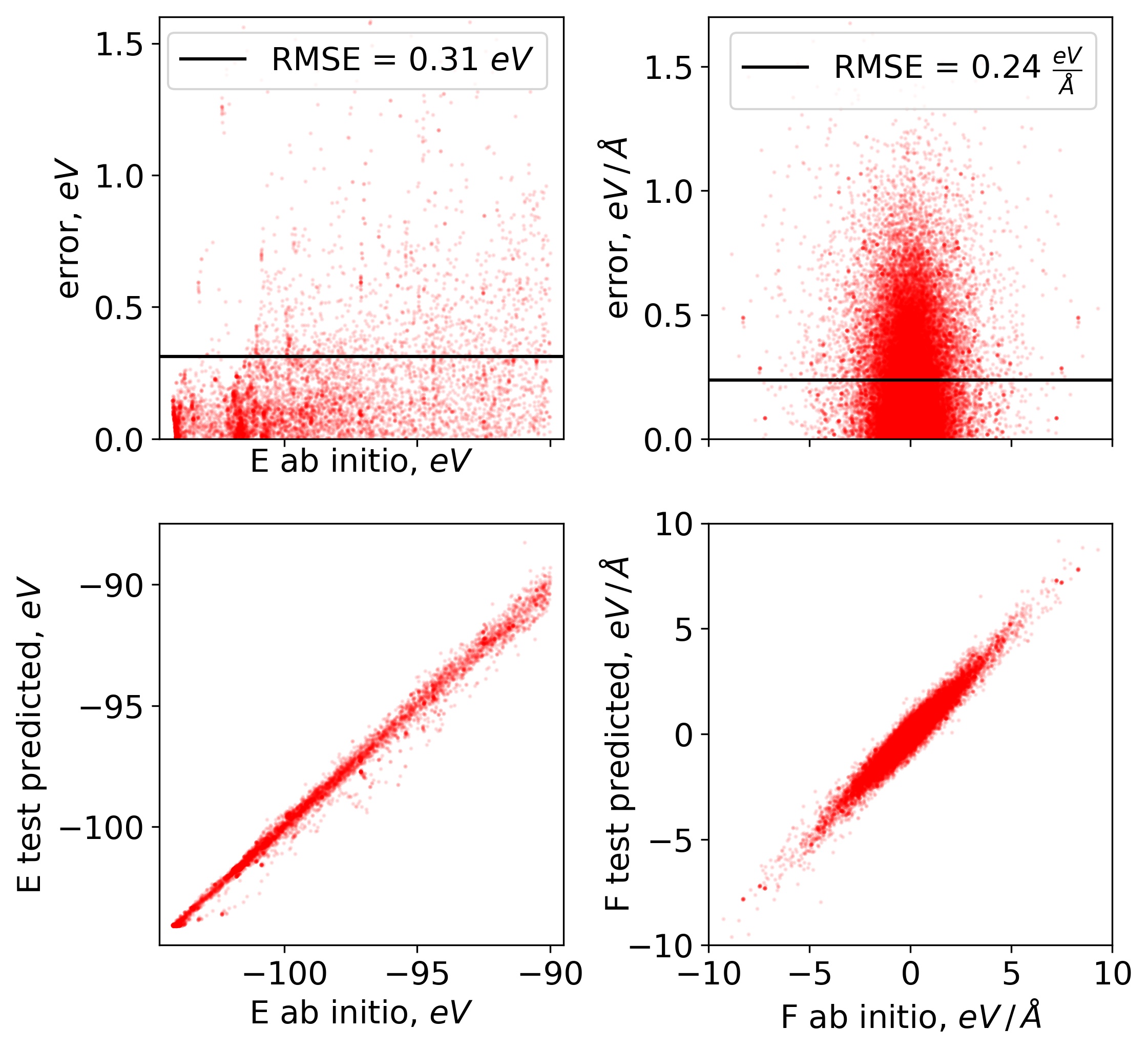}
    \caption{Performance of GTTP for tungsten.}
    \label{fig:tungsten}
\end{figure}

The intrinsic flexibility of the potential makes it transferable in terms of the types and sizes of atomic systems. The example of tungsten demonstrates the good performance of GTTP for such huge systems as grain boundaries (GBs), which are among the most challenging subjects of computational chemistry \cite{sutton1995interfaces}. For creating the potential, only the knowledge of randomly generated crystalline configurations of tungsten was used: the dataset consisted of  7286 structures with 8 atoms in the unit cell, and their energies varied from $-13.02 \frac{eV}{atom}$ to $-11.25 \frac{eV}{atom}$ with mean of $-12.35 \frac{eV}{atom}$ and standard deviation of $0.48 \frac{eV}{atom}$. The standard deviation of force components was $0.89 \frac{eV}{\si{\angstrom}}$. Values of $R^2_{cut}$ and $R^3_{cut}$ were set to 10.0 $\AA $ and 6.0 $\AA $,  respectively, and the first variant of triples pruning was chosen. The test errors of GTTP in energies and forces were $0.33 eV$ (per unit cell) or $8.5\%$ and $0.26 \frac{eV}{\si{\angstrom}}$ or $29\%$, which is illustrated in Fig. \ref{fig:tungsten}. 
 
\begin{figure*}[htbp!]
    \centering
    \includegraphics[width=1.0\textwidth]{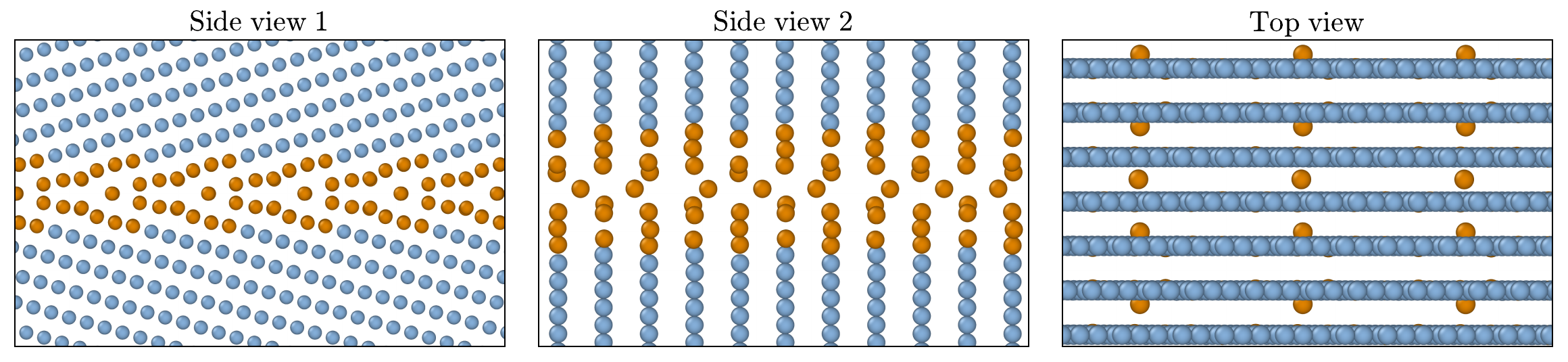}
    \caption{Ground state of the $\Sigma27[5\bar{5}2](110)$ GB from evolutionary search with GTTP, $\gamma_{\mathrm{GB}}$ = 2.55 J $\cdot$ m$^{-2}$.}
    \label{fig:gb_1007}
\end{figure*}

In order to test the performance of the constructed potential on GBs, we compared the results of grain boundaries structure prediction made using the USPEX code. In this work, a family of $\Sigma 27 (5\bar{5}2)[110]$ symmetric tilt GBs of tungsten with different atomic densities were predicted. The structures were subsequently relaxed using the LAMMPS code \cite{lammps}, employing EAM1 \cite{marinica2013} and EAM2 \cite{zhou2001atomic} potentials. In order to verify their stability, \textit{ab initio} calculations were performed. We used the same initial structures for the calculation with GTTP potential. The results of these three approaches are summarized in Table \ref{tab:potentials}. The ground state of the $\Sigma 27 (5\bar{5}2)[110]$ GB is demonstrated in Fig. \ref{fig:gb_1007}.  

\begin{table}[h!]
\begin{tabular}{ C{1.1cm}|C{1cm}|C{1.4cm}|C{1.4cm}|C{1.4cm}|C{1.4cm} }

\hline
 \textbf{Label}  & \textbf{[n]} & \textbf{EAM1} & \textbf{EAM2}  & \textbf{GTTP} & \textbf{DFT} \\
 \hline
GB1    &	1/2  &  2,819  &	   & 2,555  &	2,592\\
GB2    &	1/2  &  2,811  &	   & 2,556  &	2,593\\
GB3    &	1/2  &  2,818  &	   & 2,605  &	2,594\\
GB4    &	1/2  &  2,807  &	   & 2,606  &	2,595\\
GB5    &	1/2  &  2,817  &	   & 2,556  &	2,609\\
GB6    &	1/2  &  2,802  &	   & 2,555  &	2,610\\
GB7    &	1/2  &  2,798  &	   & 2,555  &	2,624\\
GB8    &	1/2  &  2,796  &	   & 2,555  &	2,626\\
GB9    &	1/2  &  2,812  & 	   & 2,559  &	2,628\\
GB10   &	0    &  3,171  &	   & 2,850  &	2,960\\
GB11   &	1/2  &         & 2,493 & 2.605  &	2,590\\
GB12   &	0    &         & 2,495 & 2,947  &	2,951\\
GB13   &	0    &         & 2,670 & 2,851  &	2,973\\
GB14   &	0    & 		   &       & 2,584  & 	2,680\\
\hline
\textbf{RMSE} & & 	0.203 & 0.321 & 0.065 &  \\
 \hline

\end{tabular}
\caption{Comparison of energies of $\Sigma 27 (5\bar{5}2)[110]$ symmetric tilt GBs with EAM1, EAM2, GTTP potentials and DFT. Within EAM potentials GB14 structure is unstable. Atomic density [n] is indicated in the second column. Root-mean-square error (RMSE) with respect to DFT was used as a quality metric of the algorithms. All data are given in J $\cdot$ m$^{-2}$ units.}
\label{tab:potentials}
\end{table}

\begin{figure}[h!]
    \centering
    \includegraphics[width=0.45\textwidth]{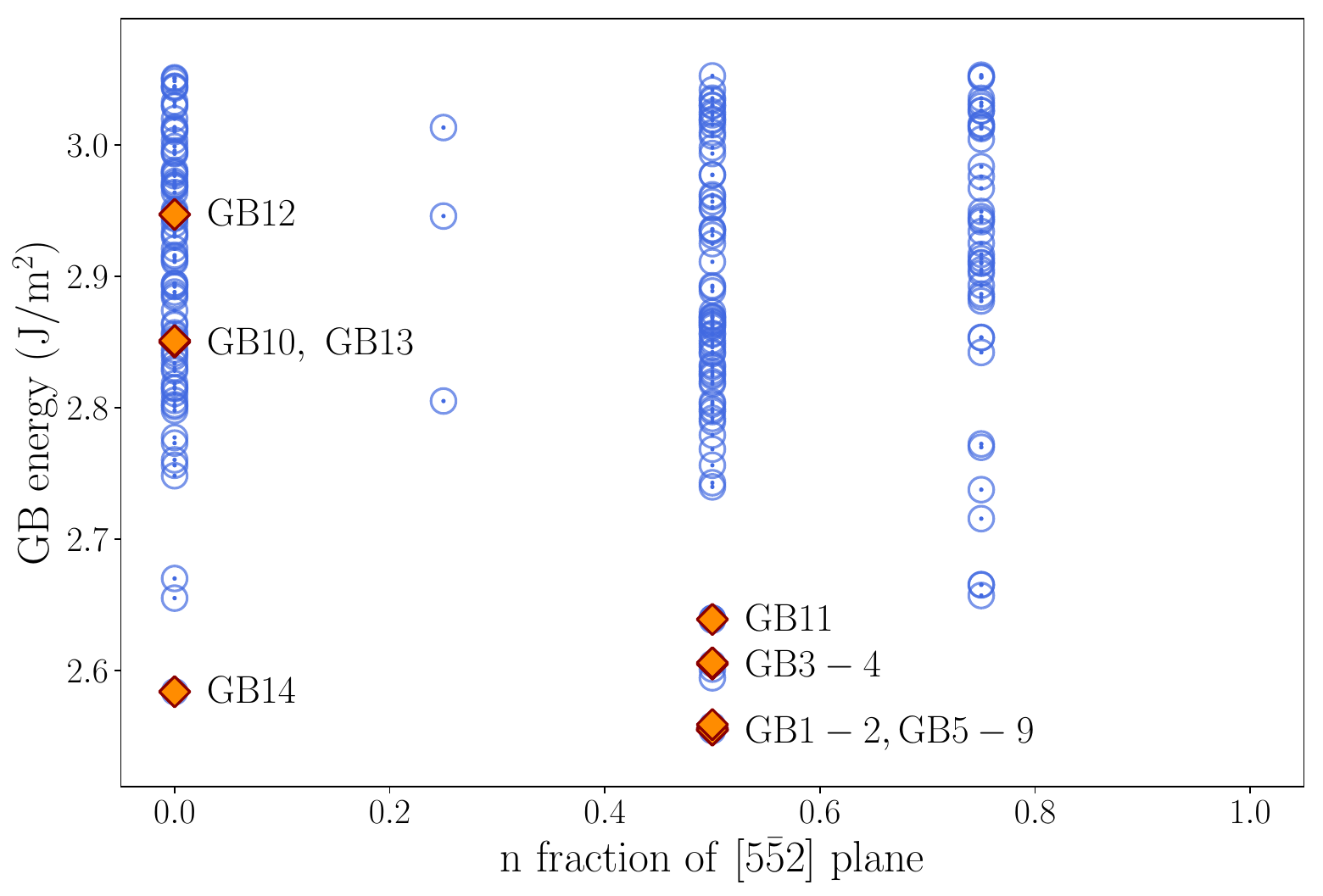}
    \caption{Results of the evolutionary search with GTTP. The GB energy is plotted as a function of atomic density [n]. GB1 -- GB14 structures from Ref. \cite{Frolov2018Nanoscale} are marked with orange diamonds. }
    \label{fig:uspex}
\end{figure}

Despite the good agreement between GTTP and DFT results, both these methods operated with the structures which were previously generated by USPEX and relaxed by EAM potentials. Therefore, we performed the same evolutionary search but used our GTTP for structure relaxation. Fig. \ref{fig:uspex} demonstrates the results of the search. Obtained GBs and their energies are marked by blue circles, while orange diamonds correspond to the most stable GBs predicted within EAM potentials \cite{Frolov2018Nanoscale}. The energy is plotted as a function of atomic density [n]. 

\begin{figure*}[ht!]
    \centering
    \includegraphics[width=0.7\textwidth]{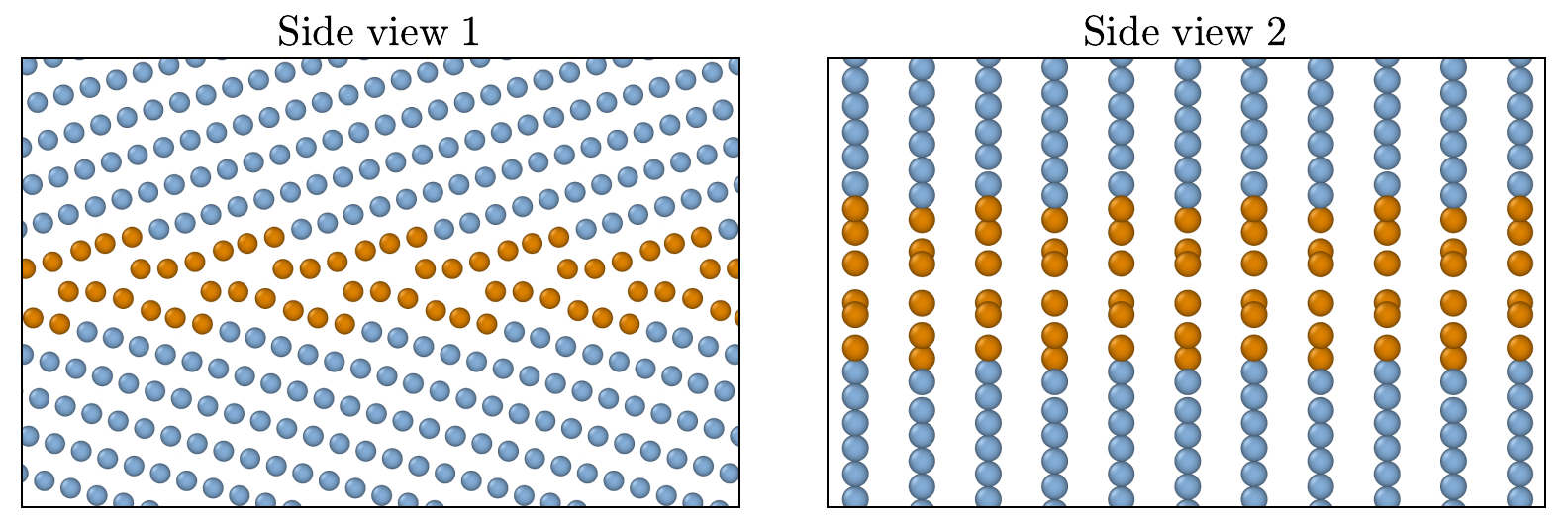}
    \caption{Metastable GB14 structure from \cite{setyawan2014},, and also predicted in this work with USPEX and GTTP, $\gamma_{\mathrm{GB}}$ = 2.58 J $\cdot$ m$^{-2}$.}
    \label{fig:gb_529}
    
\end{figure*}

Thus, all the structures from Table \ref{tab:potentials} were found by evolutionary search with GTTP. Comparison of the energy values shows that GTTP practically removes ambiguity in the ground state representation, which plagued EAM potentials, and provides 3-5 times better accuracy. It is worth noting that the metastable GB14 structure (Figure \ref{fig:gb_529}) with [n]=0, which was previously discovered in work \cite{setyawan2014}, was found by evolutionary search, while both EAM potentials treated it as an unstable one. 

\subsection{Performance on two- and three component systems}
\begin{figure}[ht!]
	\centering
	\includegraphics[width=8cm]{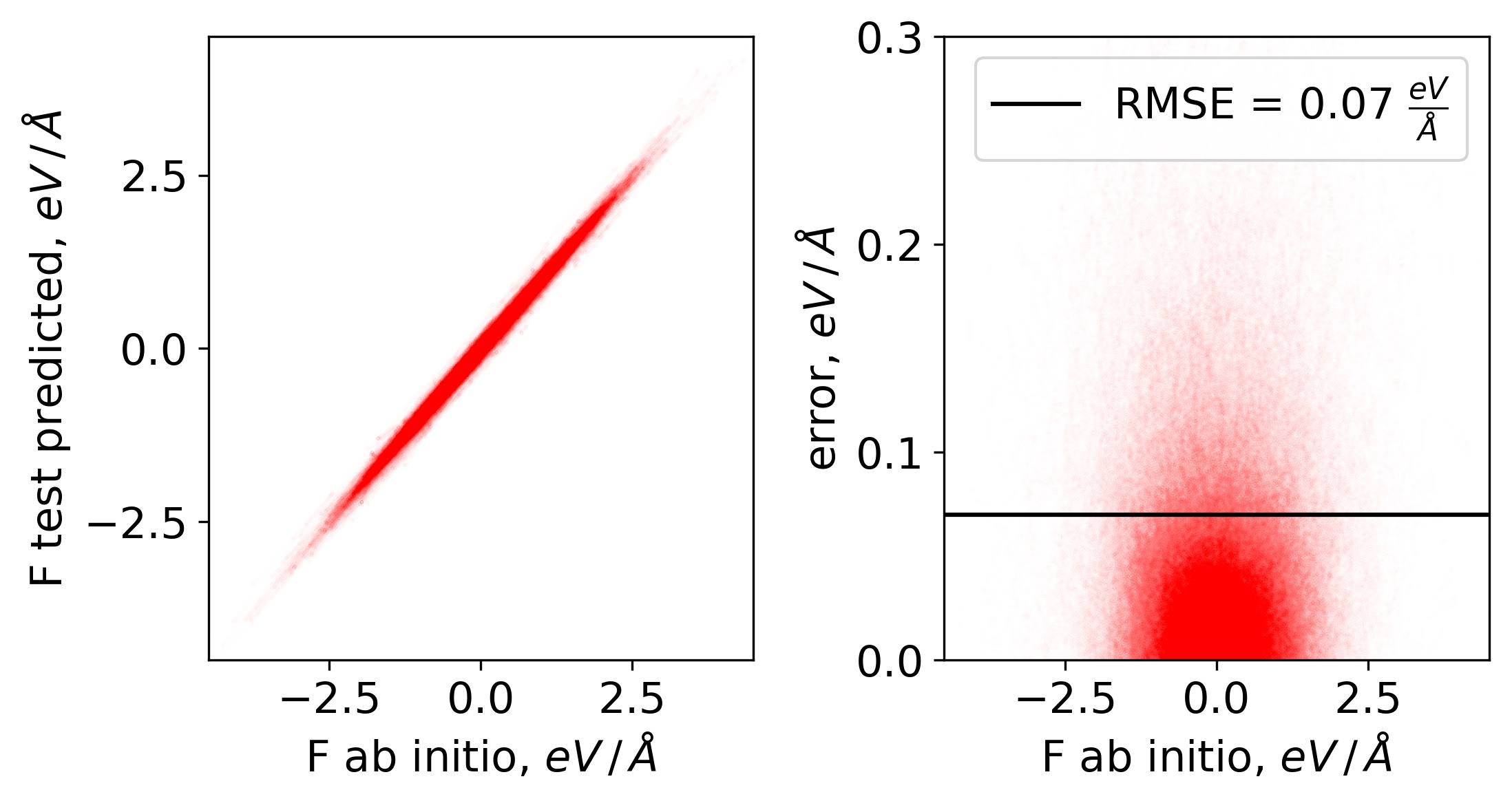}
	\caption{Performance of GTTP on titanium hydride}
	\label{fig:TiH}
\end{figure}

To test our potential on multicomponent systems 
we applied it to titanium hydride and Li-intercalated anatase $\text{Ti} \text{O}_2$. The titanium hydride dataset contained 17335 steps of \textit{ab initio} molecular dynamics trajectory with 108 titanium and 189 hydrogen atoms in the unit cell. This was taken from our recent study \cite{Arslan}. The force component standard deviation is  $0.92\:\frac{eV}{\si{\angstrom}}$. We trained our potential on the first third of the molecular dynamics trajectory and tested on the last. We choose $R^2_{\text{cut}} = \SI{10.0}{\angstrom}$,  $R^3_{\text{cut}} = \SI{4.34}{\angstrom}$ and  second variant of triples cutting. The error turned out to be  $0.070\:\frac{eV}{\si{\angstrom}}$ or $7.6\%$ , which is illustrated in Fig. \ref{fig:TiH}.

\begin{figure}[ht!]
	\centering
	\includegraphics[width=8cm]{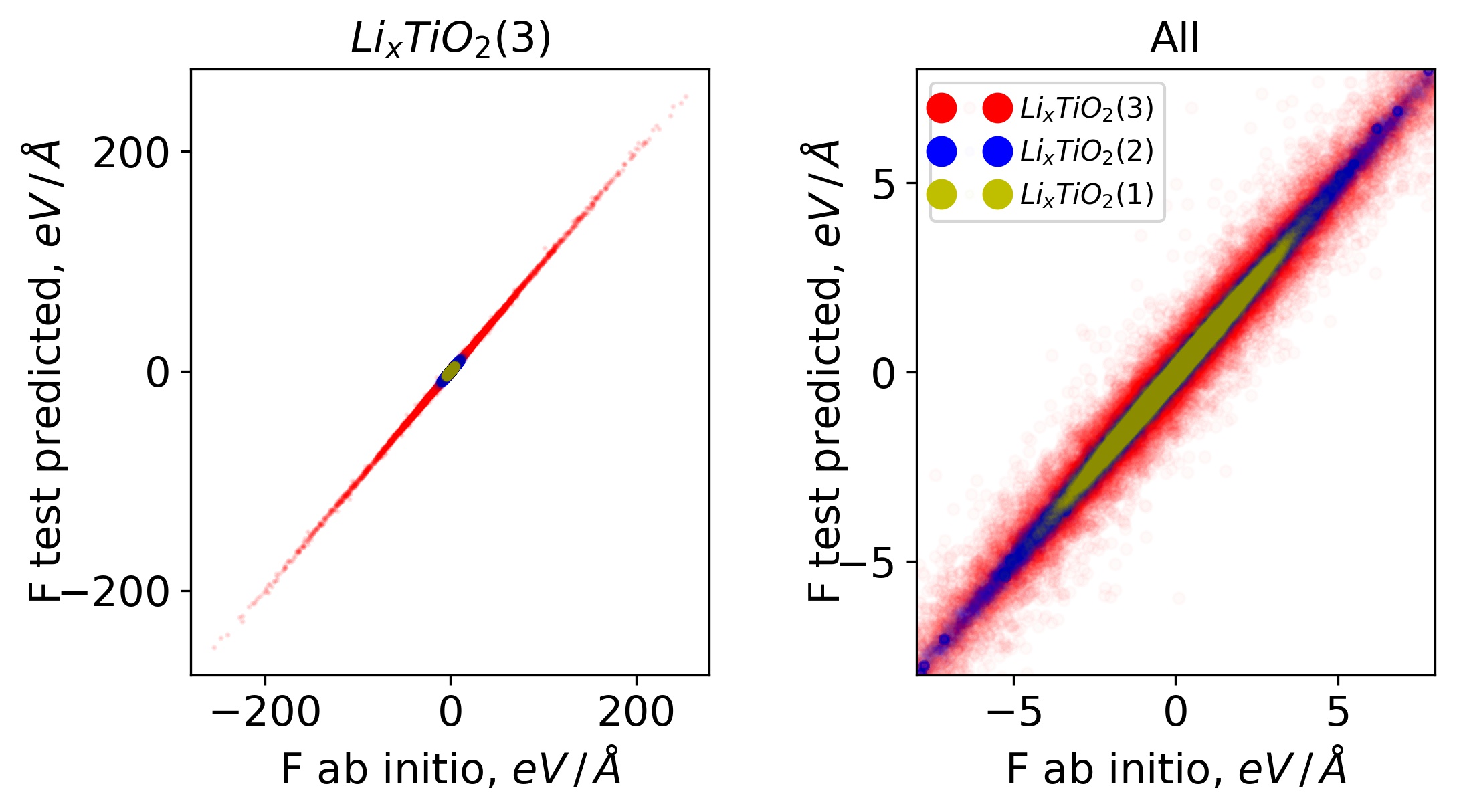}
	\caption{Performance of GTTP on Li-intercalated anatase}
	\label{fig:three_components}
\end{figure}

In case of  Li-intercalated anatase $\text{Ti} \text{O}_2$ we used  three datasets with random structures---$\text{Li}_x \text{Ti} \text{O}_2 (1)$ \cite{ShapeevLixTiO2},   $\text{Li}_x \text{Ti} \text{O}_2 (2)$  and $\text{Li}_x \text{Ti} \text{O}_2 (3)$. Datasets $\text{Li}_x \text{Ti} \text{O}_2 (2)$ and $\text{Li}_x \text{Ti} \text{O}_2 (3)$ were generated by applying some mutations to the structures from the $\text{Li}_x \text{Ti} \text{O}_2 (1)$ dataset. All datasets contain structures with 16 titanium and 32 oxygen atoms. The number of lithium atoms varied from 1 to 14 in $\text{Li}_x \text{Ti} \text{O}_2 (1)$ and $\text{Li}_x \text{Ti} \text{O}_2 (3)$, and was equal to 14 in $\text{Li}_x \text{Ti} \text{O}_2 (2)$. Chosen hyperparameters of the potential are $R^2_{\text{cut}} = \SI{10.9}{\angstrom}$,  $R^3_{\text{cut}} = \SI{4.7}{\angstrom}$.  The numerical results of the performance of the potential are given in Table \ref{table:performance_three_component} and illustrated in Fig. \ref{fig:three_components}.

\begin{table}
	\caption{Performance of the two- and three-body potential on the Li-intercalated anatase}
	\label{table:performance_three_component}
	\begin{tabular}{ | c | c | c | c |}
		\hline
		dataset & number of structures & $\sqrt{\overline{F^2}}$ $\frac{eV}{\si{\angstrom}}$ &  RMSE F $\frac{eV}{\si{\angstrom}}$ \\ \hline
		$\text{Li}_x \text{Ti} \text{O}_2 (1)$ & 618 & 0.83 & 0.086, 10.3 \%  \\ \hline
		$\text{Li}_x \text{Ti} \text{O}_2 (2)$ & 947 & 2.01 & 0.152, 7.6 \%\\ \hline
		$\text{Li}_x \text{Ti} \text{O}_2 (3)$ & 218 & 22.4 & 0.795, 3.5 \%\\ \hline
	\end{tabular}
\end{table}

The absolute error grows with the increase of standard deviations of force components or with the coverage of phase volume. But at the same time, the relative error decreases. We already faced this behavior for aluminum in Section \ref{sec:al_performance}. The same situation was also observed in \cite{deringer2017machine}.

\subsection{Chemical insights from raw data}
\begin{figure}[ht!]
	\centering
	\includegraphics[width=8cm]{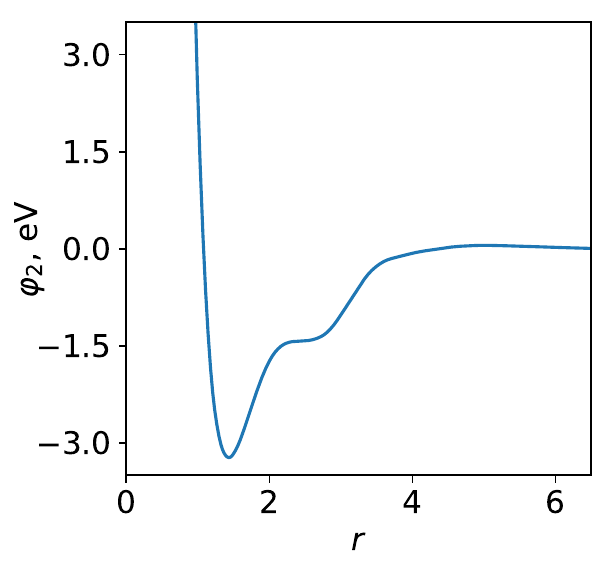}
	\caption{Two-body potential for carbon.}
	\label{fig:2_body_potential_carbon}
\end{figure}

\begin{figure}[ht!]
	\centering
	\includegraphics[width=8cm]{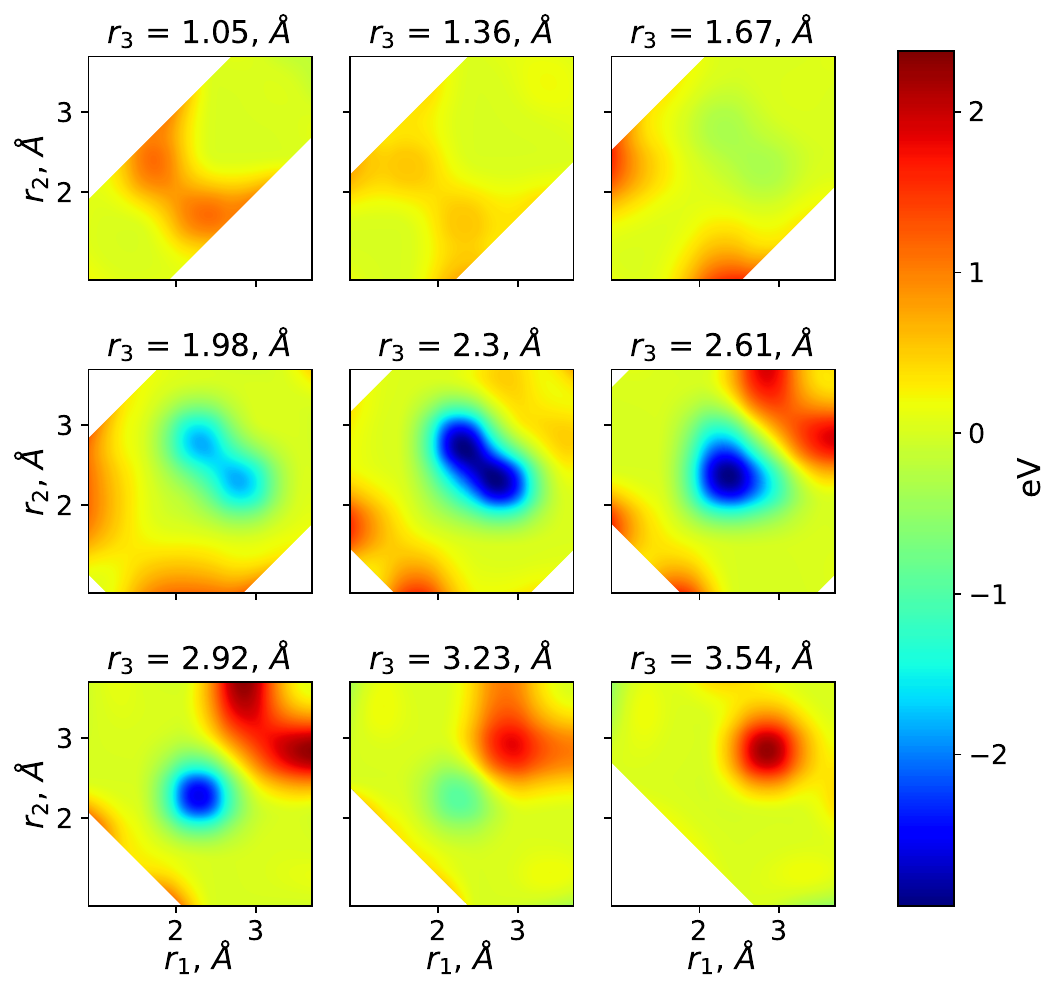}
	\caption{Three-body potential for carbon.}
	\label{fig:3_body_potential_carbon}
\end{figure}

Besides other advantages, our approach enables the extraction of interpretable information from large amounts of raw \textit{ab initio} calculations, and further, we will consider carbon as an example. In order to fit the potential, we used a dataset containing 8353 random crystal structures, each with 8 atoms in the unit cell. The energy varied from $-8.9\:\frac{eV}{atom}$  to $-5.0\:\frac{eV}{atom}$. The resulting two- and three-body potentials are shown in Fig. \ref{fig:2_body_potential_carbon} and \ref{fig:3_body_potential_carbon}.

\begin{figure}[ht!]
	\centering
	\includegraphics[width=8cm]{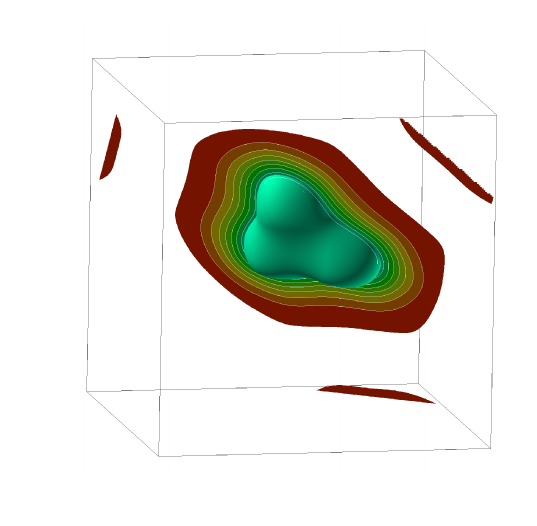}
	\caption{-2.0 eV isosurface of the 3-body potential for carbon (in the center).}
	\label{fig:isosurface_carbon}
\end{figure}

\begin{figure}[ht!]
	\centering
	\includegraphics[width=8cm]{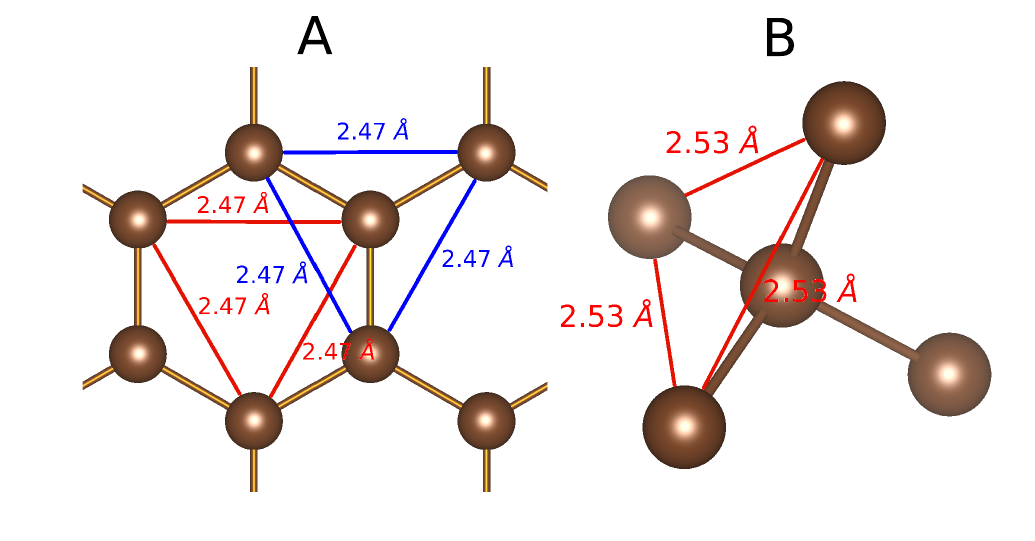}
	\caption{Equilateral C-C-C triangles in (a) graphite and (b) diamond}
	\label{fig:carbon_structures}
\end{figure}

The position of the minimum of the two-body potential is $\SI{1.43}{\angstrom}$,  which, as expected, corresponds to the C-C bond length (the C-C bond length is $\SI{1.40}{\angstrom}$ in graphite, and $\SI{1.54}{\angstrom}$ in diamond). The three-body potential has a very distinct minimum, which is also shown in the form of isosurface in Fig. \ref{fig:isosurface_carbon}, at the equilateral triangle with the side of $\SI{2.47}{\angstrom}$. This means that carbon should prefer crystal structures with such triangles. As Fig. \ref{fig:carbon_structures} \cite{momma2011vesta, Jain2013} shows, both graphite and diamond contain such equilateral triangles with the side of approximately $\SI{2.5}{\angstrom}$. 

In addition, the importance of two- and three-body interactions in various systems can be studied. In order to do it, we gathered statistics for three archetypal cases - nearly-free-electron metal (aluminum), metal with a significant directional component of bonding (tungsten), and a covalent substance (carbon), which is shown in Table \ref{table:chemical_insights}.

\begin{table*}
\centering
\caption{Importance of two- and three-body interactions in aluminum, tungsten, and carbon. For every dataset, the following information is included: 1) standard deviation of \textit{ab initio} energies in the dataset; 2) standard deviation of energies predicted by only two-body component of two- and three-body potential; 3) same for the three-body component; 4) RMSE error of only two-body potential; 5) RMSE error of two- and three-body potential; 6-10) the same for forces.} 
    \begin{tabular}{| c | c | c | c |}
    \hline
    
    dataset & aluminum $\text{Rand}_2$ & tungsten & carbon \\ \hline 
    STD E, $eV$ &  9.16 & 3.85 & 5.79\\ \hline 
    STD E 2-body, $eV$ & 9.47 & 5.70& 7.50\\ \hline 
    STD E 3-body, $eV$ & 1.30 & 3.63 & 5.80\\ \hline 
    RMSE E only 2-body, $eV$ &  0.54, 5.9\% & 0.63, 16.5\% & 4.42, 76.3\%\\ \hline 
    RMSE E 2- and 3-body, $eV$  & 0.21, 2.27\% & 0.33, 8.5\%& 2.02, 35.0 \%\\ \hline 
    STD F, $\frac{eV}{\si{\angstrom}}$ &  5.00& 0.89  &  5.81 \\ \hline 
    STD F 2 body, $\frac{eV}{\si{\angstrom}}$ & 4.92 &  0.77 & 5.64\\ \hline 
    STD F 3-body, $\frac{eV}{\si{\angstrom}}$ &  0.48 & 1.08 & 2.83\\ \hline 
    RMSE F only 2-body, $\frac{eV}{\si{\angstrom}} $ &  0.29, 5.8\% & 0.46, 51.9\% & 1.96, 33.8\%\\ \hline 
    RMSE F 2- and 3-body, $\frac{eV}{\si{\angstrom}}$ &  0.12, 2.4\% & 0.26, 29.3\%& 1.17, 20.2\%\\ \hline

    \end{tabular}
\label{table:chemical_insights}
\end{table*}
In the case of aluminum, the two-body description can reproduce most of the variability in energies and forces. The error of only two-body potential is relatively low, and, in the case of two- and three-body potential, the three-body part plays the role of small correction. The situation is the opposite for tungsten and carbon. In this case, the three-body interactions are very important, and moreover, correlations of higher order make a noticeable
contribution to the energy variance. 

\section{conclusion}
We have developed the framework for constructing two- and three-body potentials. 
Our methodology allows to model any two- and three-body interactions with arbitrary precision. At the same time, computational costs do not depend on the number of parameters or fitting flexibility and constitute a constant time per every considered pair or triple of atoms.

We applied our potential to aluminum, tungsten, titanium hydride, Li-intercalated anatase $\text{Ti} \text{O}_2$, and carbon. In the case of aluminum, it showed great accuracy and good transferability properties---we found that the potential trained on only small random structures is able to describe with satisfactory accuracy large structures from a different distribution than in the training dataset. This is even more noticeable in the case of tungsten, where we used only random structures with just 8 atoms in the unit cell as training dataset and then applied the potential to study large-scale grain boundaries in polycrystalline structures. We found that our potential significantly outperforms conventional EAM potentials specifically prepared for this purpose. In terms of RMSE of surface energy, our potential is 3--5 times better. 

We studied the tradeoff between accuracy and computational time given by the developed potential on aluminum and found that our potential has good accuracy already at the times of the order of $10^{-6} - 10^{-5}\: \frac{sec}{atom}$. 

The fitting procedure of our potential is very simple and reduces to linear regression. The number of hyperparameters is relatively small, and the influence of each of them was studied in detail. It is not necessary to search over hyperparameters for every new dataset from scratch. $Q_2$, $Q_3$ and $Im$ can be transferred directly, while $R_{\text{cut}}^2$ and $R_{\text{cut}}^3$ can be chosen in such a way as to ensure the same number of considered pairs and triples of atoms. It approximately corresponds to the same average number of neighbors within the spheres of radii $R_{\text{cut}}^2$ and $R_{\text{cut}}^3$.

In addition, the shape of the two- and three-body potential itself can provide useful chemical insights, as shown by the example of carbon. But such interpretations should be made with great care because the potential depends not only on the chemical properties of corresponding atoms but also on the distribution of structures in the training dataset, as well as on hyperparameters. 



\section {Authors contribution}
S. P., A. R. O., and E. M. designed the research, S. P. developed the interatomic potential and tested it for different systems, A. M. and I. K.  investigated GBs in tungsten. All authors wrote the manuscript.

\section{Declaration of interests}
The authors declare that they have no known competing financial interests or personal relationships that could have appeared to influence the work reported in this paper.

\section{acknowledgements}
We thank Alexandr Shapeev for sharing $\text{Li}_x \text{Ti} \text{O}_2(1)$ dataset along with constructing the Pareto front of Moment Tensor Potential on aluminum. We are very grateful to Oleg Sergeev for his invaluable assistance in integrating our potential to LAMMPS and to Aleksey Yanilkin for useful discussions. Also, we thank Qiang Zhu for implementing structure prediction of GBs. A. R. O thanks Russian Science Foundation (grant 19-72-30043). I.K. and A.M. thank the Russian Science Foundation (grant No. 21-73-10261) for financial support. 

\bibliography{bibliography}


\end{document}